\RequirePackage{fix-cm}
\documentclass[smallextended]{svjour3}       % onecolumn (second format)
\usepackage{natbib}

\smartqed  % flush right qed marks, e.g. at end of proof
\usepackage{graphicx}
\journalname{Empirical Software Engineering}
\usepackage{amsmath,amssymb,amsfonts}
\usepackage{algorithmic}
\usepackage{textcomp}
\usepackage{rotating}
\usepackage{multirow}
\usepackage{booktabs}
\usepackage{tabularx}
\usepackage{mdframed}

\newcommand{\FSCORE}{\textit{F1 score}}
\newcommand{\RECALL}{\textit{recall}}
\newcommand{\PRECISION}{\textit{precision}}
\usepackage[nolist]{acronym}
\begin{acronym}[ECU]
\acro{ADT}{Alternating Decision Trees}
\acro{BN}{Bayesian Network}
\acro{CARM}{Classification Association Rule Mining}
\acro{DT}{Decision Trees}
\acro{HDP}{Hierarchical Dirichlet Process}
\acro{LDA}{Latent Dirichlet Allocation}
\acro{LR}{Logistic Regression}
\acro{LSTM}{Long Short-Term Memory}
\acro{MNB}{Multinomial Na\"ive Bayes}
\acro{NB}{Na\"ive Bayes}
\acro{RF}{Random Forest}
\acro{SVM}{Support Vector Machine}
\acro{TFM}{Term Frequency Matrix}
\acro{IDF}{Inverse Document Frequency}
\end{acronym}

\usepackage{hyperref}

\begin{document}

\newcommand\setrow[1]{\gdef\rowmac{#1}#1\ignorespaces}
\newcommand\clearrow{\global\let\rowmac\relax}
\clearrow

\title{On the Feasibility of Automated Prediction of Bug and Non-Bug Issues}

\author{Steffen Herbold \and Alexander Trautsch \and Fabian Trautsch}

\institute{Steffen Herbold\\Institute AIFB, Karlsruhe Institute of Technology (KIT), Karlsruhe, Germany\\
           \email{steffen.herbold@kit.edu}
           \vspace{5pt}\\
           Alexander Trautsch\\Institute of Computer Science, University of Goettingen, Germany\\
           \email{alexander.trautsch@cs.uni-goettingen.de}
           \vspace{5pt}\\
           Fabian Trautsch\\Institute of Computer Science, University of Goettingen, Germany\\
           \email{fabian.trautsch@cs.uni-goettingen.de}
}

\date{Received: date / Accepted: date}

\maketitle

\begin{abstract}
Context: Issue tracking systems are used to track and describe tasks in the development process, e.g., requested feature improvements or reported bugs. However, past research has shown that the reported issue types often do not match the description of the issue. 

Objective: We want to understand the overall maturity of the state of the art of issue type prediction with the goal to predict if issues are bugs and evaluate if we can improve existing models by incorporating manually specified knowledge about issues. 

Method: We train different models for the title and description of the issue to account for the difference in structure between these fields, e.g., the length. Moreover, we manually detect issues whose description contains a null pointer exception, as these are strong indicators that issues are bugs.

Results: Our approach performs best overall, but not significantly different from an approach from the literature based on the fastText classifier from Facebook AI Research. The small improvements in prediction performance are due to structural information about the issues we used. We found that using information about the content of issues in form of null pointer exceptions is not useful. We demonstrate the usefulness of issue type prediction through the example of labelling bugfixing commits. 

Conclusions: Issue type prediction can be a useful tool if the use case allows either for a certain amount of missed bug reports or the prediction of too many issues as bug is acceptable.

\end{abstract}

\section{Introduction}

The tracking of tasks and issues is a common part of modern software engineering, e.g., through dedicated systems like Jira and Bugzilla, or integrated into other other systems like GitHub Issues. Developers and sometimes users of software file issues, e.g., to describe bugs, request improvements, organize work, or ask for feedback. This manifests in different \emph{types} into which the issues are classified. However, past research has shown that the issue types are often not correct with respect to the content~\citep{Antoniol2008,Herzig2013,Herbold2020}. 

Wrong types of issues can have different kinds of negative consequences, depending on the use of the issue tracking system. We distinguish between two important use cases, that are negatively affected by misclassifications. First, the types of issues are important for measuring the progress of projects and project planing. For example, projects may define a quality gate that specifies that all issues of type bug with a major priority must be resolved prior to a release. If a feature request is misclassified as bug this may hold up a release. Second, there are many Mining Software Repositories (MSR) approaches that rely on issue types, especially the issue type bug, e.g., for bug localization~\citep[e.g.,][]{Marcus2004, Lukins2008, Rao2011, Mills2018} or the labeling of commits as defective with the SZZ algorithm~\citep{Sliwerski2005} and the subsequent use of these labels, e.g., for defect prediction~\citep[e.g.,][]{Hall2012, Hosseini2017, Herbold2018} or the creation of fine-grained data~\citep[e.g.,][]{Just2014}. Mislabelled issues threaten the validity of the research and would also degenerate the performance of approaches based on this data that are implemented in tools and used by practitioners. Thus, mislabeled issues may have direct negative consequences on development processes as well as indirect consequences due to the downstream use of possibly noisy data. Studies by \cite{Herzig2013} and \cite{Herbold2020} have independently and on different data shown that on average about 40\% issues are mislabelled, and most mislabels are issues wrongly classified as BUG. 

There are several ways on how to deal with mislabels. For example, the mislabels could be ignored and in case of invalid blockers manually corrected by developers. Anecdotal evidence suggests that this is common in the current state of practice. Only mislabels that directly impact the development, e.g., because they are blockers, are manually corrected by developers. With this approach the impact of mislabels on the software development processes is reduced, but the mislabels may still negatively affect processes, e.g., because the amount of bugs is overestimated or because the focus is inadvertently on the addition of features instead of bug fixing. MSR would also still be affected by the mislabels, unless manual validation of the data is done, which is very time consuming~\citep{Herzig2013, Herbold2020}. 

Researchers suggested an alternative through the automated classification of issue types by analyzing the issue titles and descriptions with unsupervised machine learning based on clustering the issues~\citep{limsettho2014automatic, limsettho2014automatic, hammad2018automatic, chawla2018automated} and supervised machine learning that create classification models~\citep{Antoniol2008, Pingclasai2013, limsettho2014comparing, chawla2015automated, zhou2016combining, Terdchanakul2017, Pandey2018, qin2018classifying, zolkeply2019classifying, otoom2019automated, kallis2019ticket}. There are two possible use cases for such automated classification models. First, they could be integrated into the issue tracking system and provide recommendations to the reporter of the issue. This way, mislabeled data in the issue tracking system could potentially be prevented, which would be the ideal solution. The alternative is to leave the data in the issue tracking unchanged, but use machine learning as part of software repository mining pipelines to correct mislabeled issues. In this case, the status quo of software development would remain the same, but the validity of MSR research results and the quality of MSR tools based on the issue types would be improved. 

Within this article, we want to investigate if machine learning models for the prediction of issue types can be improved by incorporating a-priori knowledge about the problem through predefined rules. For example, null pointer exceptions are almost always associated with bugs. Thus, we investigate if separating issues that report null pointers from those that do not contain null pointers improves the outcome. Moreover, current issue type prediction approaches ignore that the title and description of issues are structurally different and simply concatenate the field for the learning. However, the title is usually shorter than the description which may lead to information from the description suppressing information from the title. We investigate if we can improve issue type prediction by accounting for this structural difference by treating the title and description separately. Additionally, we investigate if mislabels in the training data are really problematic or if they do not negatively affect the decisions made by classification models. To this aim, we compare how the training with large amounts of data that contains mislabels performs in comparison to training with a smaller amount of data that was manually validated. Finally, we address the question how mature machine learning based issue type correction is and evaluate how our proposed approach, as well as the approaches from the literature, perform in two scenarios: 1) the classification of all issues regardless of their type and 2) the classification of only issues that are reported as bug. The first scenario evaluates how good the approaches would work in recommendation systems where a label must be suggested for every incoming issue. The second scenario evaluates how good the approaches would work to correct data for MSR. We only consider the correction of issues of type bug, because both \cite{Herzig2013} and \cite{Herbold2020} found that mislabels mostly affect issues of type bug. Moreover, many MSR approaches are interested in identifying bugs. 

Thus, the research questions we address in the article are the following. 
\begin{itemize}
    \item \textbf{RQ1}: Can manually specified logical rules derived from knowledge about issues be used to improve issue type classification?
    \item \textbf{RQ2}: Does training data have to be manually validated or can a large amount of unvalidated data also lead to good classification models?
    \item \textbf{RQ3}: How good are issue type classification models at recognizing bug issues and are the results useful for practical applications?
\end{itemize}

We provide the following contributions to the state of the art through the study of these research questions.
\begin{itemize}
    \item We determined that the difference in the structure of the issue title and description may be used to slightly enhance prediction models by training separate predictors for the title and the description of issues. 
    \item We found that rules that determine supposedly easy subsets of data based on null pointers do not help to improve the quality of issue type prediction models aimed at identifying bugs. 
    \item We were successfully able to use unvalidated data to train issue type prediction models that perform well on manually validated test data with performance comparable to the currently assigned labels by developers. 
    \item We showed that issue type prediction is a useful tool for researchers interested in improving the detection of bugfixing commits. The quality of the prediction models also indicate that issue type prediction may be useful for other purposes, e.g., as recommendation system. 
    \item We provide open source implementations of the state of the art of automated issue type prediction as a Python package. 
\end{itemize}

The remainder of this paper is structured as follows. We describe the terminology we use in this paper and the problems we are analyzing in Section~\ref{sec:terminology}, followed by a summary of the related work on issue type prediction in Section~\ref{sec:related_work}. Afterwards, we discuss our proposed improvements to the state of the art in the sections~\ref{sec:approach} and~\ref{sec:noisydata}. We present the design and results of our empirical study of issue type prediction in Section~\ref{sec:experiments} and further discuss our findings in Section~\ref{sec:discussion}. Finally, we discuss the treats to the validity of our work in Section~\ref{sec:threats} before we conclude in Section~\ref{sec:conclusion}.

\section{Terminology and Problem Description}
\label{sec:terminology}

\begin{figure}
\centering
\includegraphics[width=\textwidth]{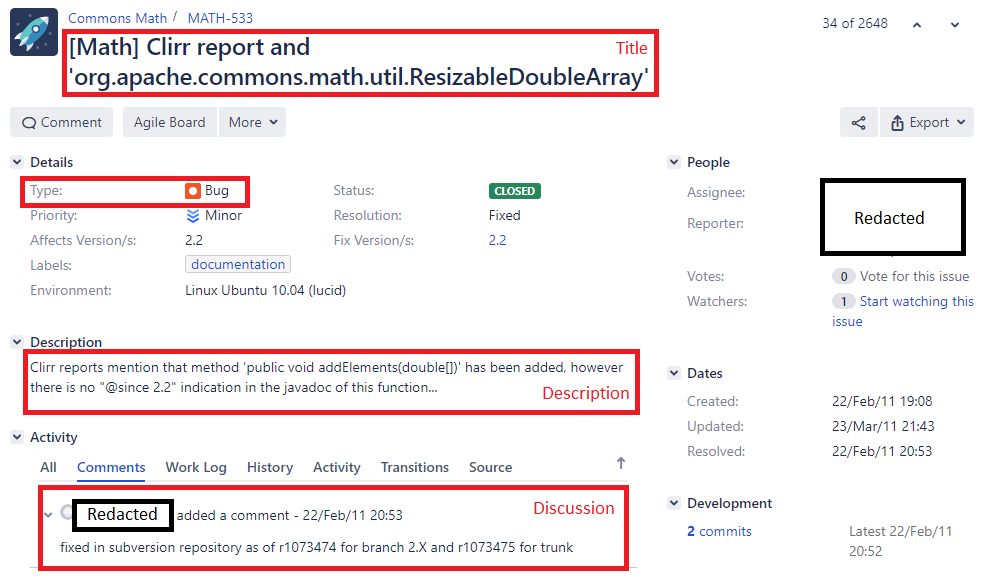}
\caption{Example of a Jira Issue from the Apache Commons Math project. Names were redacted due to data privacy concerns.}
\label{fig:jira-screen}
\end{figure}

Before we proceed with the details of the related work and our approach, we want to establish a common terminology and describe the underlying problem. Figure~\ref{fig:jira-screen} shows a screenshot of the issue MATH-533 from the Jira issue tracking system of the Apache Software Foundation. Depending on the development process and the project, issues can either be reported by anyone or just by a restricted group of users, e.g., developers, or users with paid maintenance contracts. In open source projects, it is common that everybody can report issues. Each issue contains several fields with information. For our work, the title, description, type, and discussion are relevant. The title contains a (very) brief textual summary of the issue, the description contains a longer textual description that should provide all relevant details. The reporter of an issue specifies the type and the title, although they may be edited later. The reporter of an issue also specifies the type, e.g., bug, improvement, documentation change. The concrete types that are available are usually configurable and may be project dependent. However, the type bug exists almost universally.\footnote{At least we have never seen an issue tracking system for software projects without this type.} Once the issue is reported, others can comment on the issue and, e.g., discuss potential solutions or request additional information. While the above example is for the Jira issue tracking system, similar fields can be found in other issue trackers as well, e.g., Bugzilla, Github Issues, and Gitlab Issues. 

We speak of \emph{mislabeled} issues, when the issue type does not match the description of the problem. \cite{Herzig2013} created a schema that can be used to identify the type of issues as either \emph{bug} (e.g., crashes), \emph{request for improvements} (e.g., update of a dependency), \emph{feature requests} (e.g., support for a new communication protocol), \emph{refactoring} (non-semantic change to the internal structure), \emph{documentation} (change of the documentation), or \emph{other} (e.g., changes to the build system or to the licenses). \cite{Herbold2020} used a similar schema, but merged the categories request for improvements, feature request, and refactoring into a single category called \emph{improvement} and added the category \emph{tests} (changes to tests). Figure~\ref{fig:jira-screen} shows an example for a mislabel. The reported problem is a missing Javadoc tag, i.e., the issue should be of type documentation. However, the issue is reported as bug instead. 

Since both \cite{Herzig2013} and \cite{Herbold2020} found that the main source of mislabels are issues that are reported as bug, even though they do not constitute bugs, but rather improvements of potentially sub optimal situations, we restrict our problem from a general prediction system for any type of issue to a prediction system for bugs. This is in line with the prior related work, with the exception of \cite{Antoniol2008} and \cite{kallis2019ticket}, who considered additional classes. Thus, we have a binary classification problem, with the label \emph{true} for issues that describe bugs, and \emph{false} for issues that do not describe bugs. Formally, the prediction model is a function $h_{all}: ISSUE \to \{true, false\}$, where $ISSUE$ is the space of all issues. In practice, not all information from the issue used, but instead, e.g., only the title and/or description. Depending on the scenarios we describe in the following, the information available to the prediction system may be limited. 

\begin{figure}
\centering
\includegraphics[width=0.8\textwidth]{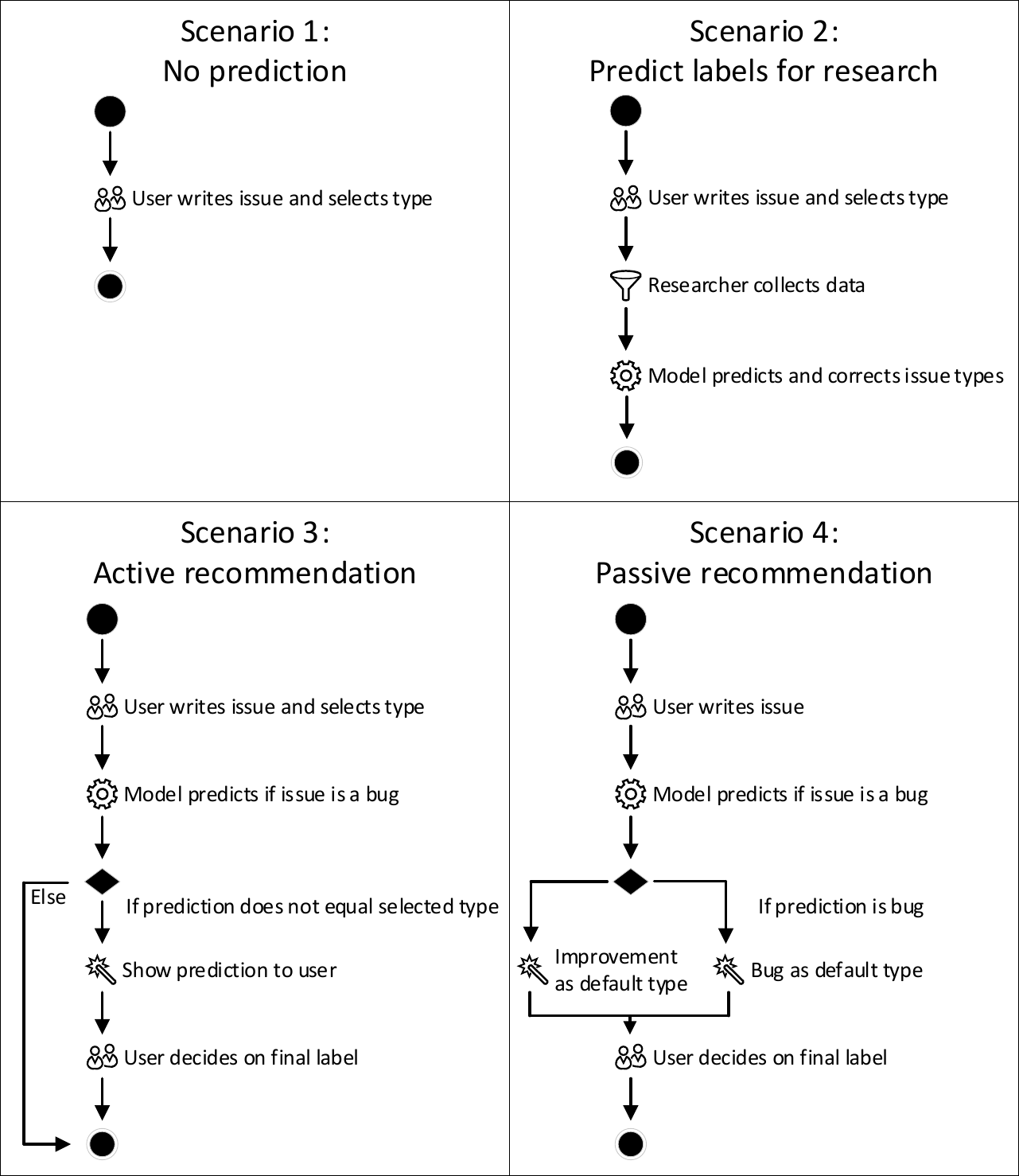}
\caption{Overview of the scenarios how prediction systems for bug issues can be used.}
\label{fig:scenarios}
\end{figure}

There are several ways such a recommendation system can be used, which we describe in Figure~\ref{fig:scenarios}. The Scenario 1 is just the status quo, i.e., a user creates an issue and selects the type. In Scenario 2, no changes are made to the actual issue tracking system. Instead, researchers use a prediction system as part of a MSR pipeline to predict issue types and, thereby, correct mislabels. In this scenario, all information from the issue tracking system is available, including changes made to the issue description, comments, and potentially even the source code change related to the issue resolution. The third and fourth scenario show how a prediction system can be integrated into an issue tracker, without taking control from the users. In Scenario 3, the prediction system gives active feedback to the users, i.e., the users decide on a label on their own and in case the prediction system detects a potential mistake, the users are asked to either confirm or change their decision. Ideally, the issue tracking system would show additional information to the user, e.g., the reason why the system thinks should be of a different type. Scenario 4 acts passively by prescribing different default values in the issue system, depending on the prediction. The rationale behind Scenario 4 is that \cite{Herzig2013} found that Bugzilla's default issue type of BUG led to more mislabels, because this was never changed by users. In Scenario 3 and Scenario 4 the information available to the prediction system is limited to the information users provide upon reporting the issue, i.e., subsequent changes or discussions may not be used. A variant of Scenario 4 would be a fully automated approach, where the label is directly assigned and the users do not have to confirm the label but would have to actively modify it afterwards. This is the approach implemented in the Ticket Tagger by \cite{kallis2019ticket}. 

Another aspect in which the scenarios differ is to which issues the prediction model is applied, depending on the goal. For example, a lot of research is interested specifically in bugs. \cite{Herzig2013} and \cite{Herbold2020} both found that almost all bugs are classified by users as type bug, i.e., there are only very few bugs that are classified otherwise in the system. To simplify the problem, one could therefore build a prediction model $h_{bug}: BUG \to \{true, false\}$ where $BUG \subset ISSUE$ are only issues which users labeled as bug. Working with such a subset may improve the prediction model for this subset, because the problem space is restricted and the model can be more specific. However, such a model would only work in Scenario 2, i.e., for use by researchers only, or Scenario 3, in case the goal is just to prevent mislabeled bugs. Scenario 4 requires a choice for all issues and would, therefore, not work with the $h_{bug}$ model. 

\section{Related Work}
\label{sec:related_work}

That classifications of issue types have a large impact on, e.g., defect prediction research was first shown by \cite{Herzig2013}. They manually validated 7,401 issue types of five projects and provided an analysis of the impact of misclassifications. They found that every third issue that is labeled as defect in the issue tracking systems is not a defect. This introduces a large bias in defect prediction models, as 39\% of files are wrongly classified as defective due to the misclassified issues that are linked to changes in the version control system. \cite{Herbold2020} independently confirmed the results by \cite{Herzig2013} and demonstrated how this and other issues negatively impact defect prediction data. However, while both \cite{Herzig2013} and \cite{Herbold2020} study the impact of mislabels of defect prediction, any software repository mining research that studies defects suffers from similar consequences, e.g., bug localization~\citep[e.g.,][]{Marcus2004, Lukins2008, Rao2011, Mills2018}.  In the literature, there are several approaches that try to address the issue of mislabels in issue systems through machine learning. These approaches can be divided into unsupervised approaches and supervised approaches. 

\subsection{Unsupervised Approaches}

The unsupervised approaches work on clustering the issues into groups and then identifying for each group their likely label. For example \cite{limsettho2014automatic,limsettho2016unsupervised} use Xmeans and EM clustering, \cite{chawla2018automated} use Fuzzy C Means clustering and \cite{hammad2018automatic} use agglomerative hierarchical clustering. However, the inherent problem of these unsupervised approaches is that they do not allow for an automated identification of the label for each cluster, i.e., the type of issue per cluster. As a consequence, these approaches are unsuited for the creation of automated recommendation systems or the use as automated heuristics to improve data and not discussed further in this article. 

\subsection{Supervised Approaches}
\label{sec:supervised}

The supervised approaches directly build classification models that predict the type of the issues. To the best of our knowledge, the first approach in this category was published by \cite{Antoniol2008}. Their approach uses the descriptions of the issues as input, which are preprocessed by tokenization, splitting of camel case characters and stemming. Afterwards, a \ac{TFM} is built including the raw term frequencies for each issue and each term. The \ac{TFM} is not directly used to describe the features used as input for the classification algorithm. Instead, \cite{Antoniol2008} first use symmetrical uncertainty attribute selection to identify relevant features. For the classification, they propose to use \ac{NB}, \ac{LR}, or \ac{ADT}. 
% They manually classified 1800 issues of three projects into issues describing a bug and others. 
% They report a precision between 64\% and 97\% and recall between 33\% and 97\% for their approach.

The \ac{TFM} is also used  by other researchers to describe the features. \cite{chawla2015automated} propose to use fuzzy logic based the \ac{TFM} on the issue title. The fuzzy logic classifier is structurally similar to a \ac{NB} classifier, but uses a slightly different scoring function. \cite{Pandey2018} propose to use the \ac{TFM} of the issue titles as input for \ac{NB}, \ac{SVM}, or \ac{LR} classifiers. \cite{otoom2019automated} propose to use a variant of the \ac{TFM} with a fixed word set. They use a list of 15 keywords related to non-bug issues (e.g., enhancement, improvement, refactoring) and calculate the term frequencies for them based on the title and description of the issue. This reduced \ac{TFM} is then used as an input for \ac{NB}, \ac{SVM}, or \ac{RF}. \cite{zolkeply2019classifying} propose to not use \ac{TFM} frequencies, but simply the occurrence of one of 60 keywords as binary features and use these to train a \ac{CARM}. \cite{Terdchanakul2017} propose to go beyond the \ac{TFM} and instead use the \ac{IDF} of n-grams for the title and descriptions of the issues as input for either \ac{LR} or \ac{RF} as classifier. 

\cite{zhou2016combining} propose an approach that combines the \ac{TFM} from the issue title with structured information about the issue, e.g., the priority and the severity. The titles are classified into the categories high (can be clearly classified as bug), low (can be clearly classified as non-bug), and middle (hard to decide) and they use the \ac{TFM} to train a \ac{NB} classifier for these categories. The outcome of the \ac{NB} is then combined with the structural information as features used to train a \ac{BN} for the binary classification into bug or not a bug. 

There are also approaches that do not rely on the \ac{TFM}. \cite{Pingclasai2013} published an approach based on topic modeling via the \ac{LDA}. Their approach uses the title, description, and discussion of the issues, preprocesses them, and calculates the topic-membership vectors via \ac{LDA} as features. \cite{Pingclasai2013} propose to use either \ac{DT}, \ac{NB}, or \ac{LR} as classification algorithm. 
% The authors evaluated their approach on three projects from the dataset by \cite{Herzig2013} and report an f-measure between 0.66 and 0.82, depending on the project and used classifier.
\cite{limsettho2014comparing} propose a similar approach to derive features via topic-modeling. They propose to use \ac{LDA} or \ac{HDP} on the title, description, and discussion of the issues to calculate the topic-membership vectors as features. For the classification, they propose to use \ac{ADT}, \ac{NB}, or \ac{LR}. In their case study, they have shown that LDA is superior to HDP. We note that the approaches by \cite{Pingclasai2013} and \cite{limsettho2014automatic} both cannot be used for recommendation systems, because the discussion is not available at the time of reporting the issue. \cite{qin2018classifying} propose to use word embeddings of the title and description of the issue as features and use these to train a \ac{LSTM}. \cite{Palacio2019} propose to use the SecureReqNet (shallow)\footnote{The network is still a deep neural network, the (shallow) means that this is the less deep variant that was used in by \cite{Palacio2019}, because they found that this performs better. neural network based on work by \cite{Han2017} for the labeling of issues as vulnerabilities. The neural network uses word embeddings and a convolutional layer that performs 1-gram, 3-gram, and 5-gram convolutions that are then combined using max-pooling and a fully connected layer to determine the classification.}

\cite{kallis2019ticket} created the tool Ticket Tagger that can be directly integrated into GitHub as a recommendation system for issue type classification. The Ticket Tagger uses the fastText~\cite{fasttext} algorithm, which uses the text as input and internally calculates a feature representation that is based on n-grams, but not of the words, but of the letters within the words. These feature are used to train a neural network for the text classification. 

The above approaches all rely on fairly common text processing pipelines to define features, i.e., the \ac{TFM}, n-grams, \ac{IDF}, topic modeling or word embeddings to derive numerical features from the textual data as input for various classifiers. In general, our proposed approach is in line with the related work, i.e., we also either rely on a standard text processing pipeline based on the \ac{TFM} and \ac{IDF} as input for common classification models or use the fastText algorithm which directly combines the different aspects. The approaches by \cite{otoom2019automated} and \cite{zolkeply2019classifying} try to incorporate manually specified knowledge into the learning process through curated keyword lists. Our approach to incorporate knowledge is a bit different, because we rather rely on rules that specify different training data sets and do not restrict the feature space.

\section{Approach}
\label{sec:approach}

Within this section, we describe our approach for issue type prediction that allows us to use simple rules to incorporate knowledge about the structure of issues and the issues types in the learning process in order to study RQ1. 

\subsection{Title and Description}
\label{sec:title-desc}

\begin{figure}
\centering
\begin{minipage}{0.20\textwidth}
\includegraphics[width=\textwidth]{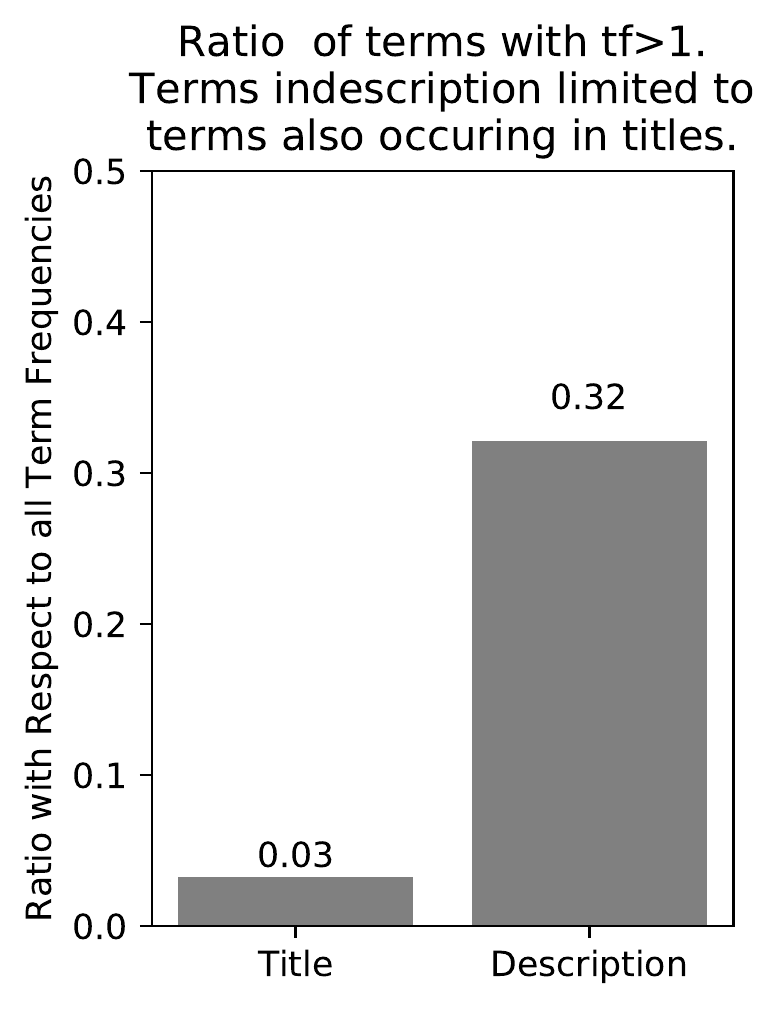}
\end{minipage}
\begin{minipage}{0.39\textwidth}
\includegraphics[width=\textwidth]{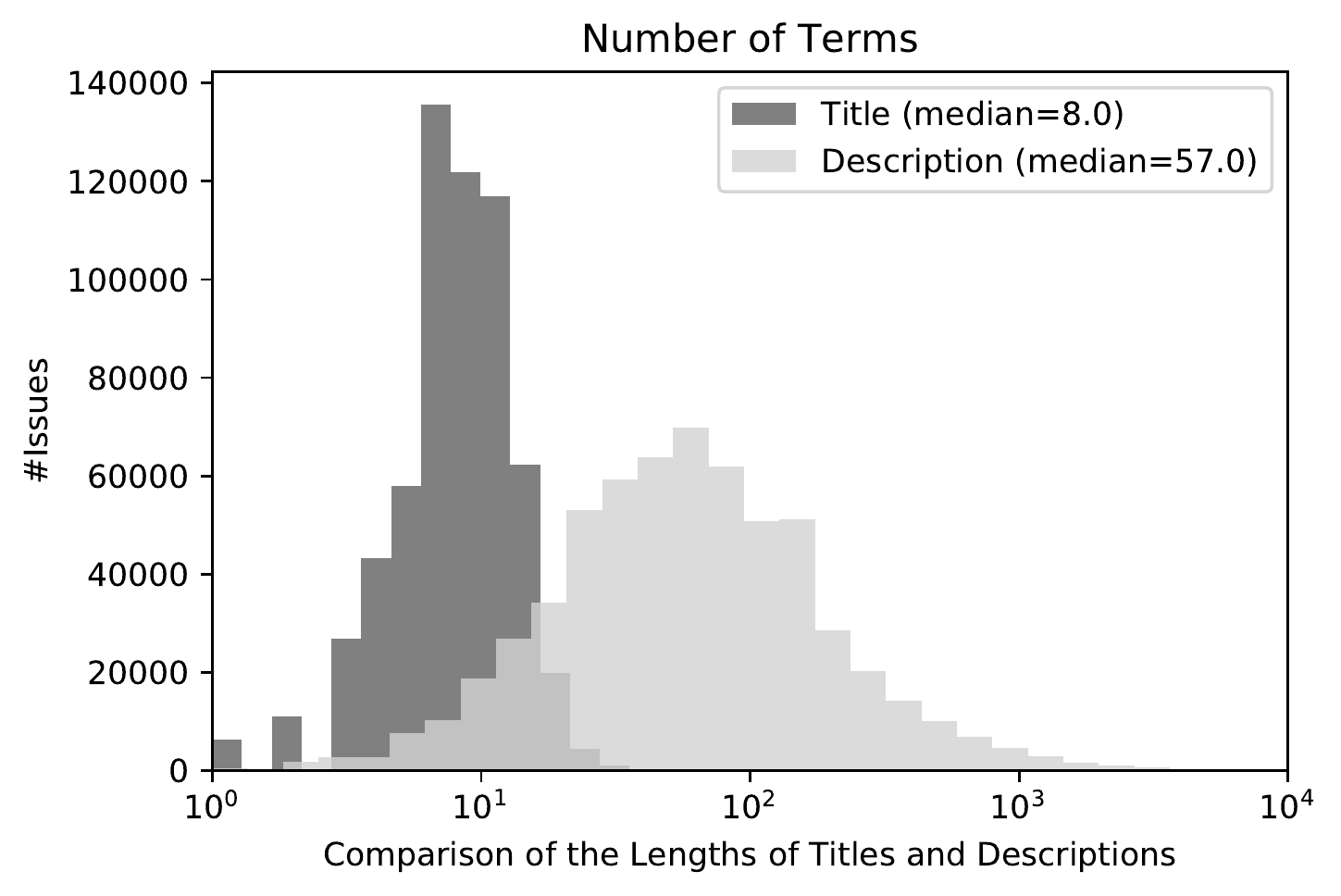}
\end{minipage}
\begin{minipage}{0.39\textwidth}
\includegraphics[width=\textwidth]{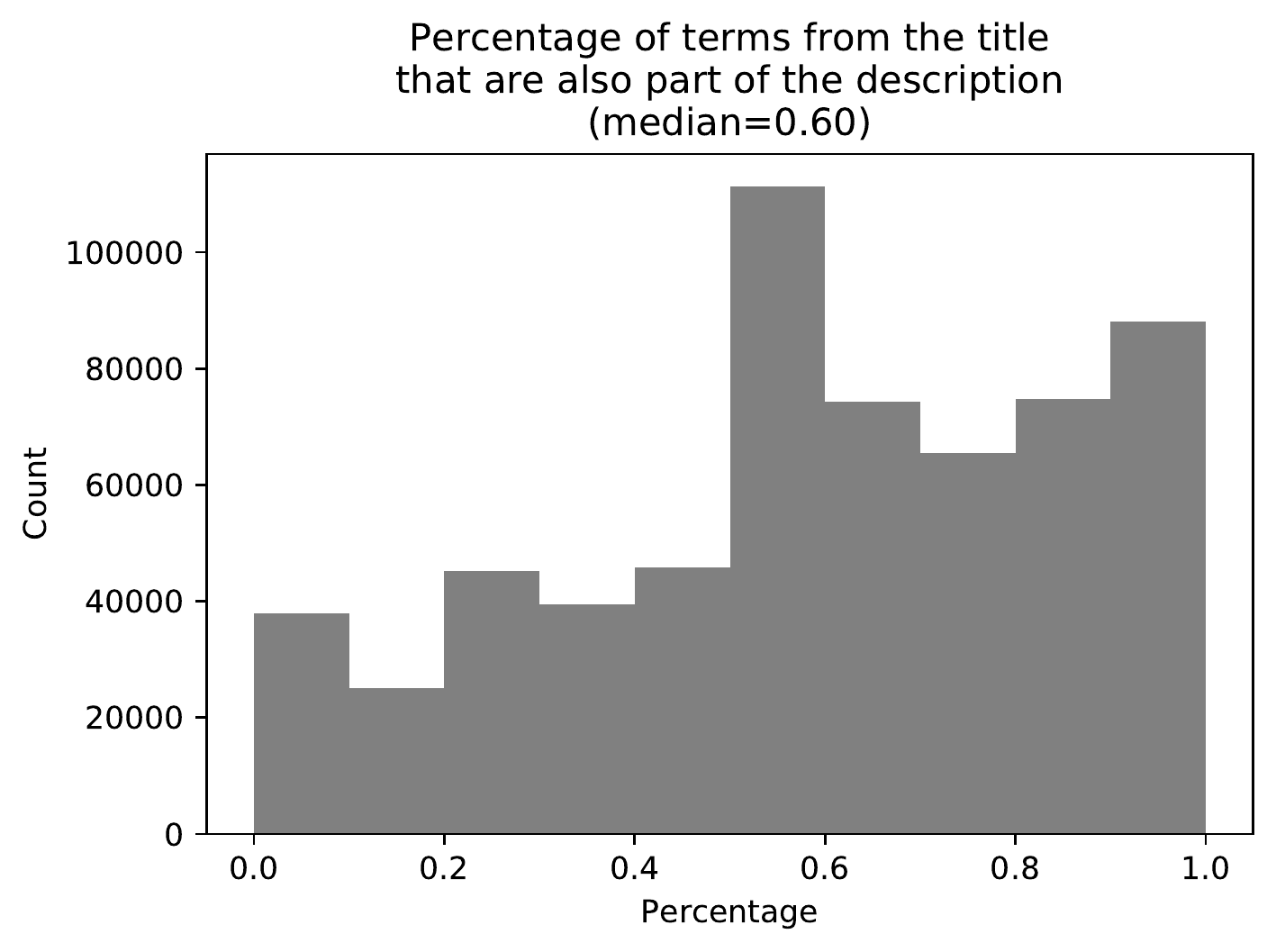}
\end{minipage}
\caption{Visualization of the structural differences of issue titles and descriptions based on the 607,636 Jira Issues from the UNVALIDATED data (see Section~\ref{sec:data}).}
\label{fig:title-vs-descr}
\end{figure}

We noticed that in the related work, researchers used the title and description together, i.e., as a single document in which the title and description are just concatenated. From our point of view, this ignores the properties of the fields, most notably the length of the descriptions. Figure~\ref{fig:title-vs-descr} shows data for the comparison of title and description. The title field is more succinct, there are almost never duplicate words, i.e., term frequency will almost always be zero or one. Moreover, the titles are very short. Thus, the occurrence of a term in a title is more specific for the issue than the occurrence of a term in the description. The description on the other hand is more verbose, and may even contain lengthy code fragments or stack traces. Therefore, many terms occur multiple times and the occurrence of terms is less specific. This is further highlighted by the overlap of terms between title and description. Most terms from the title occur also in the description and the terms lose their uniqueness when the title and description are considered together. As a result, merging of the title and description field may lead to suppressing information from the title in favor of information from the description, just due to the overall lengths of the fields and the higher term frequencies.  This loss of information due to the structure of the features is undesirable. 

We propose a very simple solution to this problem, i.e., the training of different prediction models for title and description. The results of both models can then be combined into a single result. Specifically, we suggest that classifiers that provide scores, e.g., probabilities for classes are used and the mean value of the scores is then used as prediction. Thus, we have two classifiers $h_{title}$ and $h_{description}$ that both predict values in $[0,1]$ and our overall probability that the issue is a bug is $\frac{h_{title}+h_{description}}{2}$. This generic approach works with any classifier, and could, also be extended with a third classifier for the discussion of the issues. 

\subsection{Easy subsets}
\label{sec:subsets}

\begin{figure}
\centering
\begin{minipage}{0.49\textwidth}
\includegraphics[width=\textwidth]{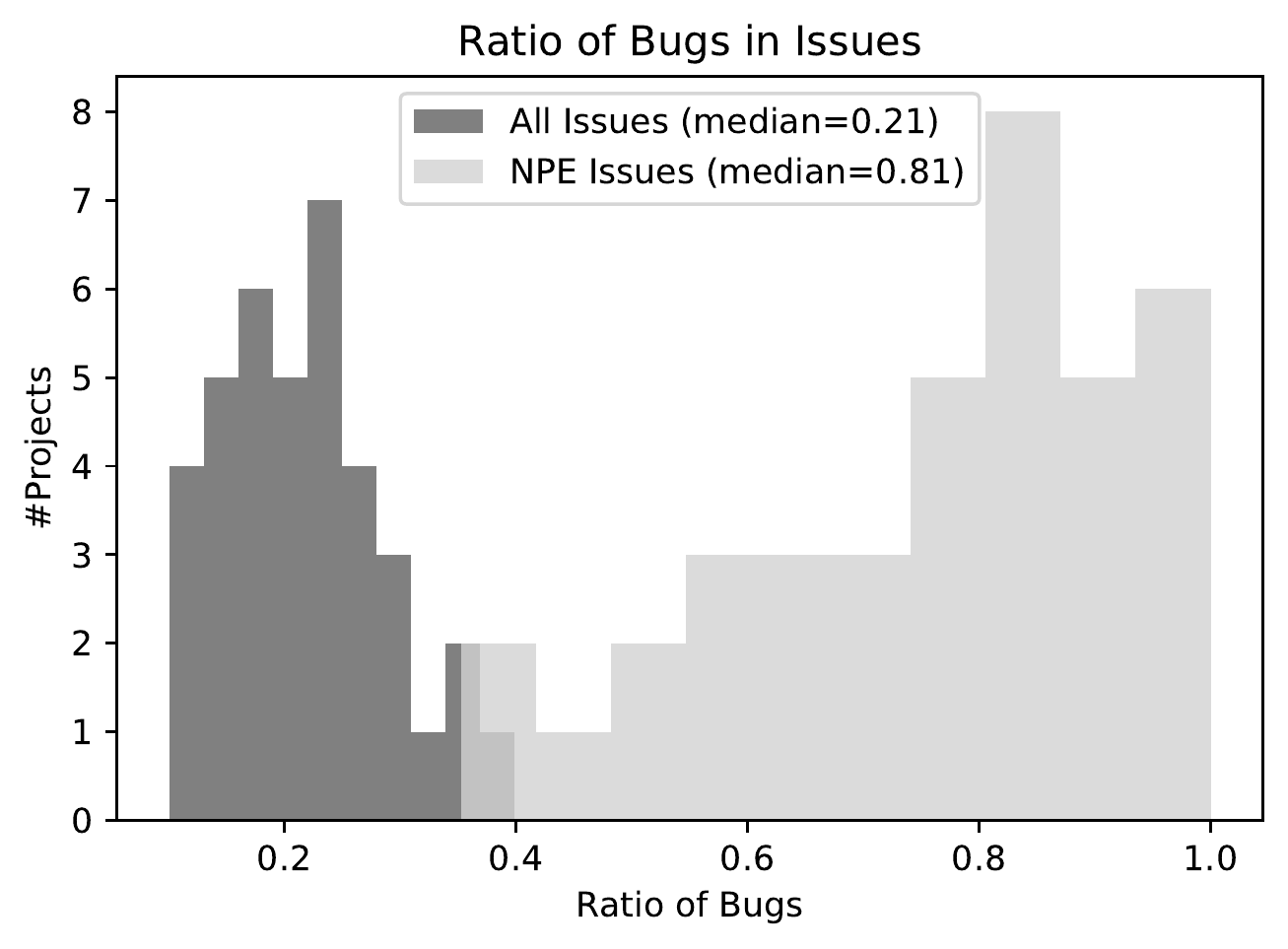}
\end{minipage}
\begin{minipage}{0.49\textwidth}
\includegraphics[width=\textwidth]{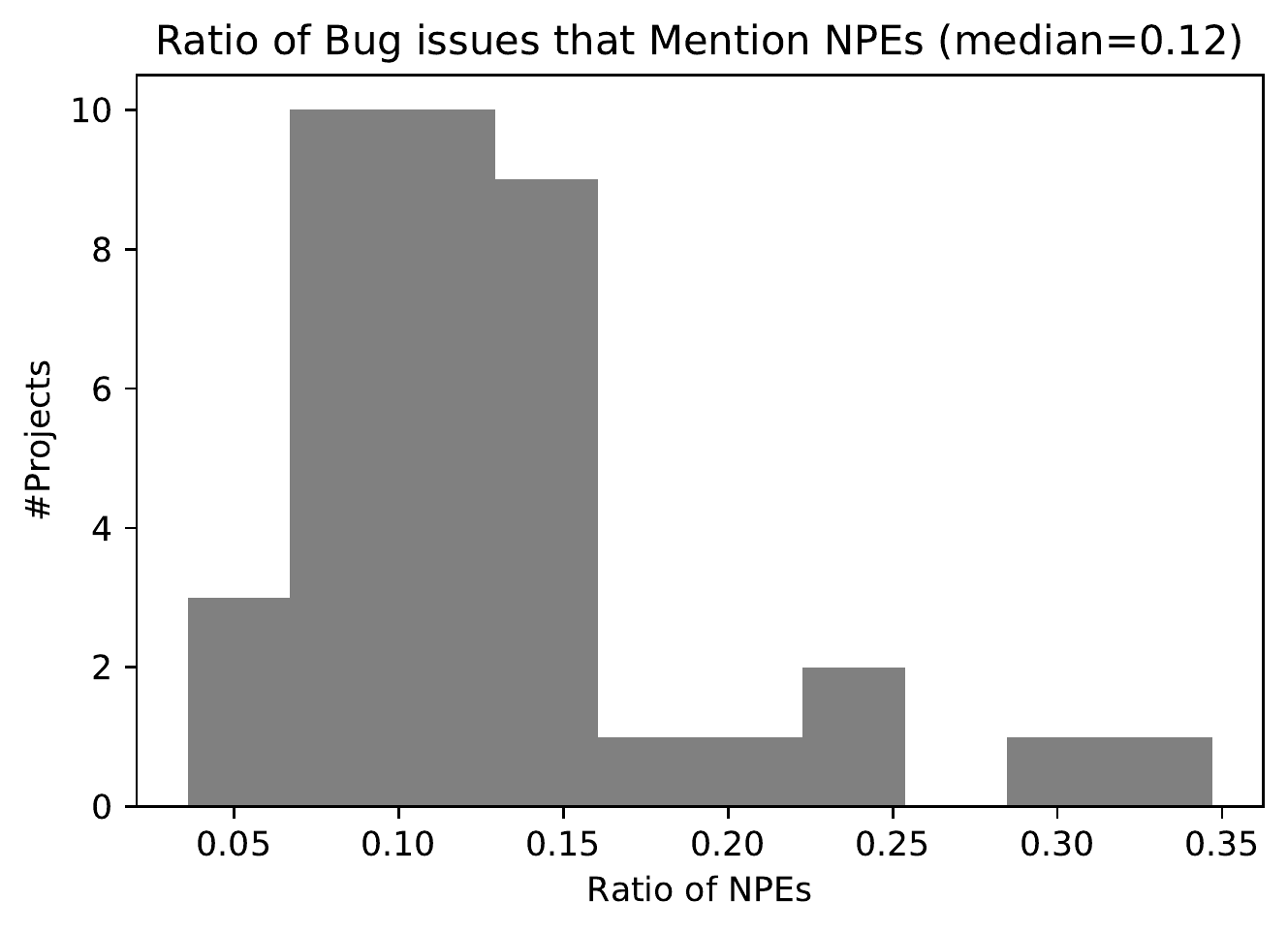}
\end{minipage}
\caption{Data on the usage of the terms NullPointerException, NPE, and NullPointer in issues based on 30,922 issues from the CV data (see Section~\ref{sec:data}).}
\label{fig:npe-assumptions}
\end{figure}

An important aspect we noted during our manual validation of bugs for our prior work \citep{Herbold2020} was that not all bugs are equal, because some rules from \cite{Herzig2013} are pretty clear. The most obvious example is that almost anything related to an unwanted null pointer exception is a bug. Figure~\ref{fig:npe-assumptions} shows that mentioning a null pointer is a good indicator for a bug and that there is a non-trivial ratio of issues that mention null pointers. However, the data also shows that a null pointer is no sure indication that the issue is actually a bug and cannot be used as a static rule. Instead, we wanted to know if we can enhance the machine learning models by giving them the advantage of knowing that something is related to a null pointer. We used the same approach as above. We decided to train one classifier for all issues that mention the terms "NullPointerException", "NPE", or "NullPointer" in the title or description, and a second classifier for all other issues. Together with the separate classifiers for title and description, we now have four classifiers, i.e., one classifier for the title of null pointer issues, one classifier for the description of null pointer issues, one classifier for the title of the other issues, and one classifier for the description of the other issues. We do not just use the average of these four classifiers. Instead, the prediction model checks if an issue mentions a null pointer and then uses either the classifiers for null pointer issues or for the other issues. 

\subsection{Classification Model}
\label{sec:classification-model}

So far, we only described that we want to train different classifiers to incorporate knowledge about issues into the learning process. However, we have not yet discussed the classifiers we propose. We consider two approaches that are both in line with the current state of the art in issue type prediction (see Section~\ref{sec:related_work}). 

The first approach is a simple text processing pipeline as can be found in online tutorials on text mining\footnote{e.g., https://www.hackerearth.com/de/practice/machine-learning/advanced-techniques/text-mining-feature-engineering-r/tutorial/\\https://scikit-learn.org/stable/tutorial/text\_analytics/working\_with\_text\_data.html} and is similar to the \ac{TFM} based approaches from the literature. As features, we use the TF-IDF of the terms in the documents. This approach is related to the \ac{TFM} but uses the \ac{IDF} as scaling factor. The \ac{IDF} is based on the number of issues in which a term occurs, i.e., 
\begin{equation}
    IDF(t) = \log\frac{n}{df(t)}+1
\end{equation}
where $n$ is the number of issues and $df(t)$ is the number of issues in which the term $t$ occurs. The TF-IDF of a term $t$ in an issue $d$ is computed as
\begin{equation}
    TF-IDF(t, d) = TF(t, d) \cdot IDF(t)
\end{equation}
where $TF(t, d)$ is the term frequency of $t$ in $d$. The idea behind using TF-IDF instead of just TF is that terms that occur in many documents may be less informative and are, therefore, down-scaled by the IDF. We use the TF-IDF of the terms in the issues as features for our first approach and use multinomial \ac{NB} and \ac{RF} as classification models. We use the TF-IDF implementation from Scikit-Learn~\citep{Pedregosa2011} with default parameters, i.e., we use the lower-case version of all terms without additional processing.

Our second approach is even simpler, taking pattern from \cite{kallis2019ticket}. We just use the fastText algorithm~\citep{fasttext} that supposedly does state of the art text mining on its own, and just takes the data as is. The idea behind this is that we just rely on the expertise of one of the most prominent text mining teams, instead of defining any own text processing pipeline. We apply the fastText algorithm once with the same parameters as were used by \cite{kallis2019ticket} and once with an automated parameter tuning that was recently made available for fastText\footnote{https://ai.facebook.com/blog/fasttext-blog-post-open-source-in-brief/}. The automated parameter tuning does not perform a grid search, but instead uses a guided randomized strategy for the hyper parameter optimization. A fixed amount of time is used to bound this search. We found that 90 seconds was sufficient for our data, but other data sets may require longer time. In the following, we refer to fastText as FT and the autotuned fastText as FTA.  

Please note that we do not consider any deep learning based text mining techniques (e.g., BERT by \cite{Devlin2018}) for the creation of a classifier, because we believe that we do not have enough (validated) data to train a deep neural network. We actually have empirical evidence for this, as the deep neural networks we used in our experiments do not perform well (see Section~\ref{sec:experiments}, Qin2018-LSTM, Palacio2019-SRN). Deep learning should be re-considered for this purpose once the requirements on data are met, e.g., through pre-trained word embeddings based on all issues reported at GitHub.

\subsection{Putting it all Together}
\label{sec:putting-it-together}

\begin{figure}
\centering
\includegraphics[width=\textwidth]{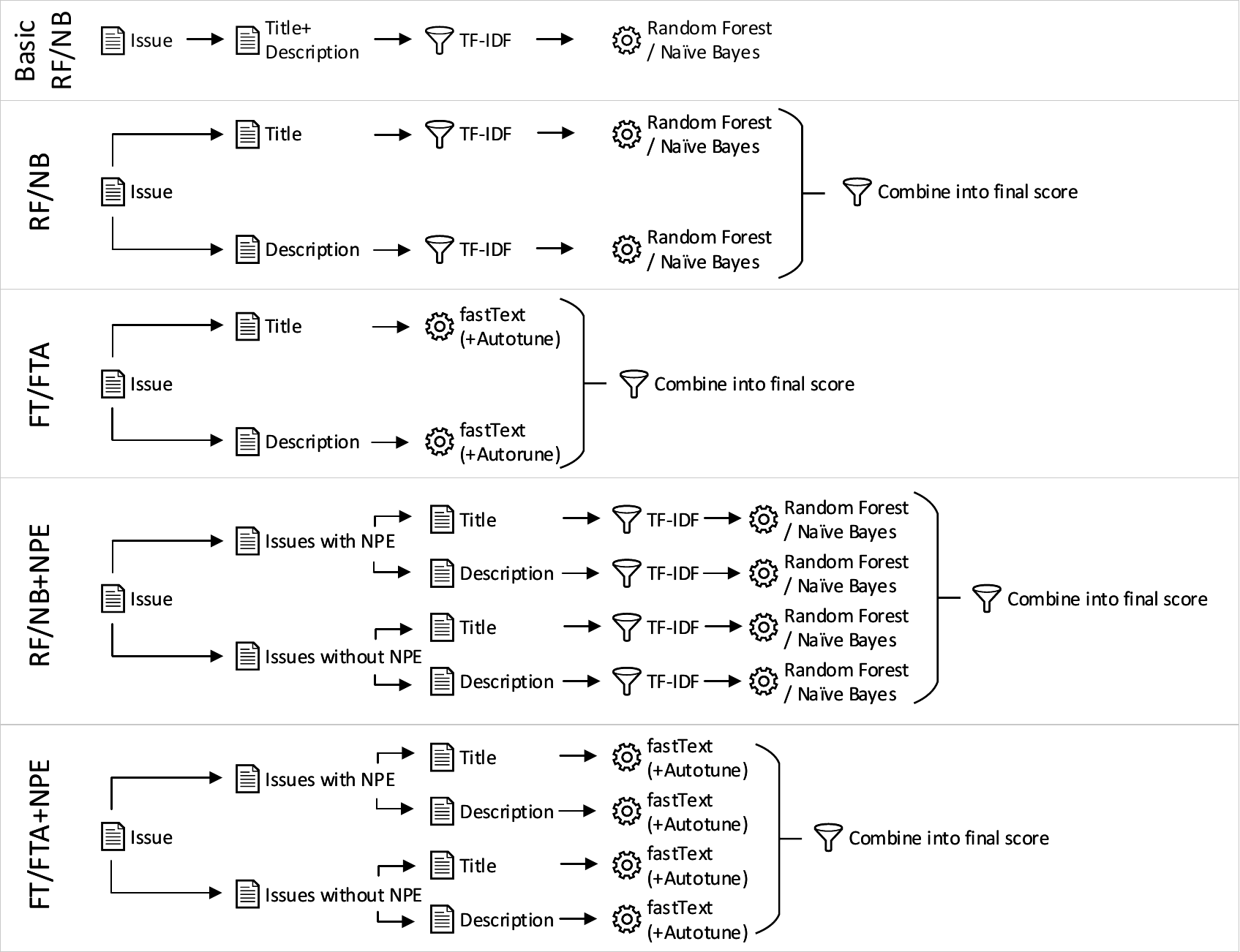}
\caption{Summary of our approach.}
\label{fig:our-approach}
\end{figure}

From the different combinations of rules and classifiers, we get ten different classification models for our approach that we want to evaluate, which we summarize in Figure~\ref{fig:our-approach}. First, we have Basic-RF and Basic-NB, which train classifiers on the merged title and description, i.e., a basic text processing approach without any additional knowledge about the issues provided by us.\footnote{Basic-FT is omitted, because this is the same as the work by \cite{kallis2019ticket} and, therefore, already covered by the literature and in our experiments in Section~\ref{sec:experiments}.} This baselines allows us to estimate if our rules actually have a positive effect over not using any rules. Next, we have RF, NB, FT, and FTA which train different classifiers for the title and description as described in Section~\ref{sec:title-desc}. Finally, we extend this with separate classifiers for null pointers and have the models RF+NPE, NB+NPE, FT+NPE, and FTA+NPE. 

\section{Unvalidated Data}
\label{sec:noisydata}

A critical issue with any machine learning approach is the amount of data is available for the training. The validated data about the issue types that accounts for mislabels is limited, i.e., there are only the data sets by \cite{Herzig2013} and \cite{Herbold2020}. Combined, they contain validated data about roughly 15,000 bugs. While this may be sufficient to train a good issue prediction model with machine learning, the likelihood of getting a good model that generalizes to many issues increases with more data. However, it is unrealistic that vast amounts of manually labelled data become available, because of the large amount of manual effort involved. The alternative is to use data that was not manually labelled, but instead use the user classification for the training. In this case, all issues from more or less any issue tracker can be used as training data. Thus, the amount of data available is huge. However, the problem is that the resulting models may not be very good, because the training data contains mislabels that were not manually corrected. This is the same as noise in the training data. While this may be a problem, it depends on where the mislabels are, and also on how much data there is that is correctly labelled. 

\begin{figure}
\centering
\includegraphics[width=\textwidth]{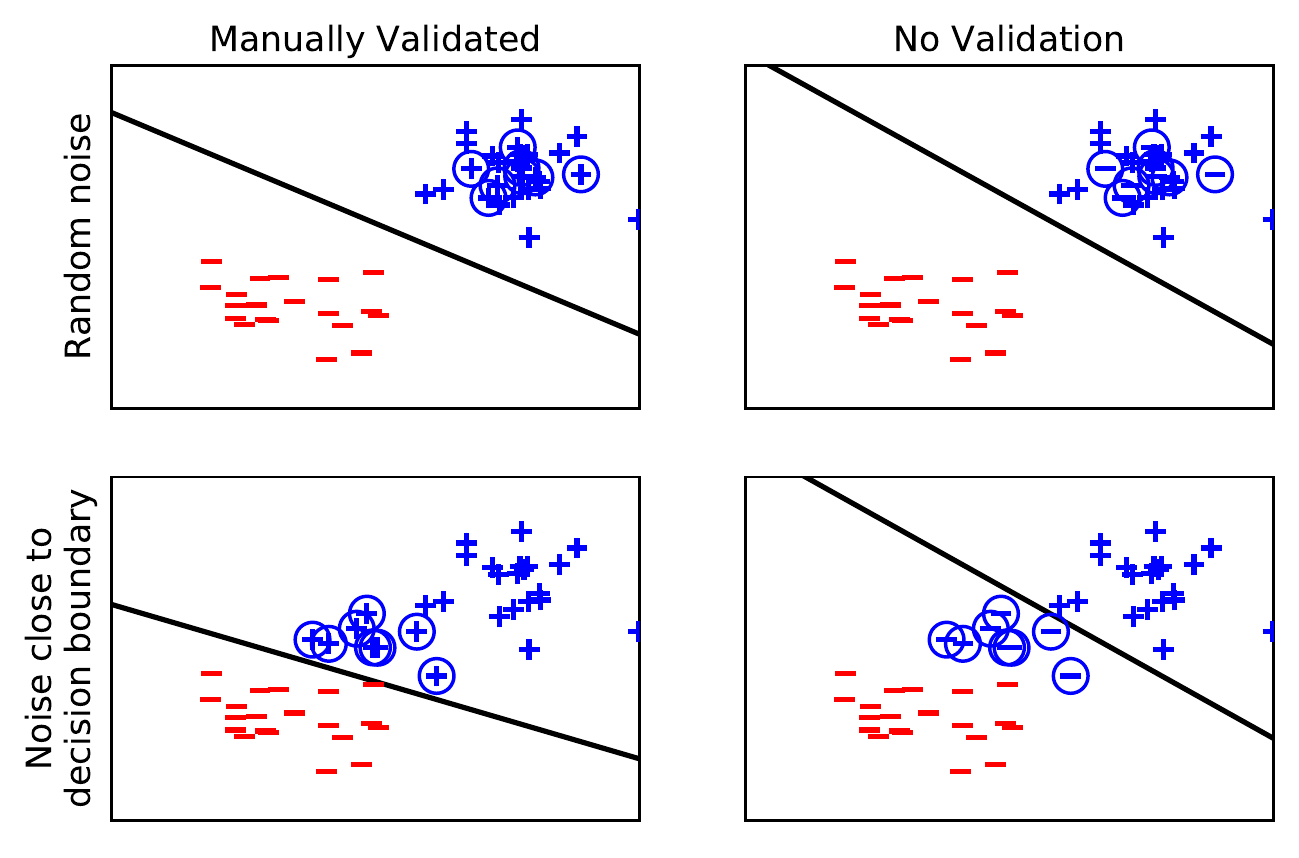}
\caption{Example for the possible effect of mislabels in the training data on prediction models. The color indicates the correct labels, i.e., red for bugs and blue for other issues. The marker indicates the label in the training data, - for bugs, + for other issues. Circled instances are manually corrected on the left side and mislabels on the right side. The line indicates the decision boundary of the classifier. Everything below the line is predicted as a bug, everything above the line is predicted as not a bug.}
\label{fig:noise}
\end{figure}

Figure~\ref{fig:noise} shows an example that demonstrates why training with unvalidated data may work and why it may fail. The first column shows data that was manually corrected, the second column shows data that was not corrected and contains mislabels. In the first row, the mislabels are random, i.e., random issues that are not a bug are mislabeled as bugs. In this case, there is almost no effect on the training, as long as there are more correctly labelled instances than noisy instances. Even better, the prediction model will even predict the noisy instances correctly, i.e., the prediction would actually be better than the labels of the training data. Thus, noise as in the first example can be ignored for training the classifier. This is line with learning theory, e.g., established by \cite{Kearns1998} who demonstrated with the statistical query model that learning in the presence of noise is possible, if the noise is randomly distributed. In the second row, the mislabels are not random, but close to the decision boundary, i.e., the issues that are most similar to bugs are mislabeled as bugs. In this case, the decision boundary is affected by the noise and would be moved to the top-right of the area without manual validation. Consequently, the trained model would still mislabel all instances that are mislabeled in the training data. In this case, the noise would lead to a performance degradation of the training and cannot be ignored.

Our hope is that mislabeled issues are mostly of the first kind, i.e., randomly distributed honest mistakes. In this case, a classifier trained with larger amounts of unlabeled data should perform similar or possibly even better than a classifier trained with a smaller amount of validated data.

\section{Experiments}
\label{sec:experiments}

We now describe the experiments we conducted to analyze issue type prediction. The experiments are based on the Python library \texttt{icb}\footnote{https://github.com/smartshark/icb} that we created as part of our work. \texttt{icb} provides implementations for the complete state of the art of supervised issue type prediction (Section~\ref{sec:supervised}) with the exceptions described in Section~\ref{sec:baselines}. The additional code to conduct our experiments is provided as a replication package\footnote{https://doi.org/10.5281/zenodo.3994254}.

\subsection{Data}
\label{sec:data}

We use four data sets to conduct our experiments. Table~\ref{tbl:data} lists statistics about the data sets. First, we use the data by \cite{Herzig2013}. This data contains manually validated data for 7,297 issues from five projects. Three projects used Jira as issue tracker (httpcomponents-client, jackrabbit, lucene-solr), the other two used Bugzilla as issue tracker (rhino, tomcat). The data shared by \cite{Herzig2013} only contains the issue IDs and the correct labels. We collected the issue titles, descriptions, and discussions for these issues with SmartSHARK~\citep{Trautsch2018, Trautsch2020}. The primary purpose of the data by \cite{Herzig2013} in our experiments is the use as test data. Therefore, we refer to this data set in the following as TEST. 

Second, we use the data by \cite{Herbold2020}. This data contains manually validated data for all 11,154 bugs of 38 projects. Issues that are not bugs were not manually validated. However, \cite{Herbold2020} confirmed the result by \cite{Herzig2013} using sampling that only about 1\% of issues that are not labeled as bugs are actually bugs. Consequently, \cite{Herbold2020} decided to ignore this small amount of noise, which we also do in this article, i.e., we assume that everything that is not labeled as bug in the data by \cite{Herbold2020} is not a bug. The primary purpose of the data by \cite{Herbold2020} in our experiments is the use in a leave-one-project-out cross-validation experiment. Therefore, we refer to this data as CV in the following. 

The third data set was collected by \cite{Ortu2015}. This data set contains 701,002 Jira issues of 1,238 projects. However, no manual validation of the issue types is available for the data by \cite{Ortu2015}. We drop all issues that have no description and all issues of projects that are also included in the data by \cite{Herzig2013} or \cite{Herbold2020}. This leaves us with 607,636 issues of 1,198 projects. Since we use this data to evaluate the impact of not validating data, we refer to this data as UNVALIDATED in the following. 

We use two variants of the data by \cite{Herzig2013} and \cite{Herbold2020}: 1) only the issues that were labelled as bug in the issue tracker; and 2) all issues regardless of their type. Our rationale for this are the different possible use cases for issue type prediction, we outlined in Section~\ref{sec:terminology}. Using these different sets, we evaluate how good issue type prediction works in different circumstances. With the first variant, we evaluate how good the issue type prediction models work for the correction of mislabeled bugs either as recommendation system or by researchers. With the second variant we evaluate how good the models are as general recommendation systems. We refer to these variants as TEST$_{BUG}$, CV$_{BUG}$,TEST$_{ALL}$, and CV$_{ALL}$. We note that such a distinction is only possible with data that was manually validated, hence, there is no such distinction for the UNVALIDATED data. 

We also use a combination of the UNVALIDATED and the CV data. The latest issue in the UNVALIDATED data was reported on 2014-01-06. We extend this data with all issues from the CV data that were reported prior to this date. We use the original labels from the Jira instead of the manually validated labels from \cite{Herbold2020}, i.e., an unvalidated version of this data that is cut off at the same time as the UNVALIDATED data. We refer to this data as UNVALIDATED+CV. Similarly, we use a subset of the CV$_{ALL}$ data, that only consists of the issues that were reported after 2014-01-06. Since we will use this data for evaluation, we use the manually validated labels by \cite{Herbold2020}. We drop the commons-digester project from this data, because only nine issues were reported after 2014-01-06, none of which were bugs. We refer to this data as CV$_{2014+}$.

Finally, we also use data about validated bugfixing commits. The data we are using also comes from \cite{Herbold2020}, who in addition to the validation of issue types also validated the links between commits and issues. They found that the main source of mislabels for bug fixing commits are mislabeled issue types, i.e., bugs that are not actually bugs. We use the validated links and validated bug fix labels from \cite{Herbold2020}. Since the projects are the same as for the CV data, we list the data about the number of bug fixing commits per project in Table~\ref{tbl:data} together with the CV data, but refer to this data in the following as BUGFIXES. 

\begin{table}
\centering
\begin{tabular}{lrrrrr}
\toprule
{} &  \multicolumn{3}{c}{Issues} & \multicolumn{2}{c}{Bugfixing Commits}\\
& All & Dev. Bugs &  Val. Bugs & No Val. & Val. \\
\midrule
httpcomponents-client &   744 &        468 &             304 & - & -\\
jackrabbit            &  2344 &       1198 &             925 & - & -\\
lucene-solr           &  2399 &       1023 &             688 & - & -\\
rhino                 &   584 &        500 &             302 & - & -\\
tomcat                &  1226 &       1077 &             672 & - & -\\
\midrule
TEST Total            &  7297 &       4266 &            2891 & - & -\\
\midrule\midrule
ant-ivy               &  1168 &        544 &             425 & 708 & 568 \\
archiva               &  1121 &        504 &             296 & 940 & 543\\
calcite               &  1432 &        830 &             393 & 923 & 427 \\
cayenne               &  1714 &        543 &             379 & 1272 & 850 \\
commons-bcel          &   127 &         58 &              36 & 85 & 49 \\
commons-beanutils     &   276 &         88 &              51 & 118 & 59 \\
commons-codec         &   183 &         67 &              32 & 137 & 59 \\
commons-collections   &   425 &        122 &              49 & 180 & 88 \\
commons-compress      &   376 &        182 &             124 & 291 & 206 \\
commons-configuration &   482 &        193 &             139 & 340 & 243 \\
commons-dbcp          &   296 &        131 &              71 & 191 & 106 \\
commons-digester      &    97 &         26 &              17 & 38 & 26 \\
commons-io            &   428 &        133 &              75 & 216 & 129 \\
commons-jcs           &   133 &         72 &              53 & 104 & 72 \\
commons-jexl          &   233 &         87 &              58 & 239 & 161 \\
commons-lang          &  1074 &        342 &             159 & 521 & 242 \\
commons-math          &  1170 &        430 &             242 & 721 & 396 \\
commons-net           &   377 &        183 &             135 & 235 & 176 \\
commons-scxml         &   234 &         71 &              47 & 123 & 67 \\
commons-validator     &   265 &         78 &              59 & 101 & 73 \\
commons-vfs           &   414 &        161 &              92 & 195 & 113 \\
deltaspike            &   915 &        279 &             134 & 490 & 217 \\
eagle                 &   851 &        230 &             125 & 248 & 130 \\
giraph                &   955 &        318 &             129 & 360 & 141 \\
gora                  &   472 &        112 &              56 & 208 & 99 \\
jspwiki               &   682 &        288 &             180 & 370 & 233 \\
knox                  &  1125 &        532 &             214 & 860 & 348 \\
kylin                 &  2022 &        698 &             464 & 1971 & 1264 \\
lens                  &   945 &        332 &             192 & 497 & 276 \\
mahout                &  1669 &        499 &             241 & 710 & 328 \\
manifoldcf            &  1396 &        641 &             310 & 1340 & 671 \\
nutch                 &  2001 &        641 &             356 & 976 & 549 \\
opennlp               &  1015 &        208 &             102 & 353 & 144 \\
parquet-mr            &   746 &        176 &              81 & 241 & 120 \\
santuario-java        &   203 &         85 &              52 & 144 & 95 \\
systemml              &  1452 &        395 &             241 & 583 & 304 \\
tika                  &  1915 &        633 &             370 & 1118 & 670 \\
wss4j                 &   533 &        242 &             154 & 392 & 244 \\
\midrule
CV Total        & 30922 &      11154 &            6333 & 18539 & 10486 \\
\midrule
UNVALIDATED Total & 607636 & 346621 & - & - & -\\ 
\bottomrule
\end{tabular}
\caption{Statistics about the data we used, i.e., the number of issues in the projects (All), the number of issues that developers labeled as bug (Dev. Bugs), the number of issues that are validated as bugs (Val. Bugs), the number of bugfixing commits without issue type validation (No Val.) and the number of bugfixing commits with issue type validation (Val.). The statistics for the BUGFIXES data set are shown in the last two columns of the CV data.}
\label{tbl:data}
\end{table}

\subsection{Baselines}
\label{sec:baselines}

Within our experiments, we not only evaluate our own approach which we discussed in Section~\ref{sec:approach}, but also compare our work to several baselines. First, we use a trivial baseline which assumes that all issues are bugs. Second, we use the approaches from the literature as comparison. We implemented the approaches as they were described and refer to them by the family name of the first author, year of publication, and acronym for the classifier. The approaches from the literature we consider are (in alphabetatical order) Kallis2019-FT by \cite{kallis2019ticket}, Palacio-2019-SRN by \cite{Palacio2019}, Pandey2018-LR and Pandey2018-NB by \cite{Pandey2018}, Qin2018-LSTM by \cite{qin2018classifying}, Otoom2019-SVC, Otoom2019-NB, and Otoom2019-RF by \cite{otoom2019automated}, Pingclasai2013-LR and Pingclasai2013-NB by \cite{Pingclasai2013}, Limsettho2014-LR and Limsettho2014-NB by \cite{limsettho2014comparing}, and Terdchanakul2017-LR and Terdchanakul2017-RF by \cite{Terdchanakul2017}. 

We note that this is, unfortunately, a subset of the related work discussed in Section~\ref{sec:related_work}. We omitted all unsupervised approaches, because they require manual interaction to determine the issue type for the determined clusters. The other supervised approaches were omitted due to different reasons. \cite{Antoniol2008} perform feature selection based on a \ac{TFM} by pair-wise comparisons of all features. In comparison to \cite{Antoniol2008}, we used data sets with more issues which increased the number of distinct terms in the \ac{TFM}. As a result, the quadratic growth of the runtime complexity required for the proposed feature selection did not terminate, even after waiting several days. \cite{zhou2016combining} could not be used, because their approach is based on different assumptions on the training data, i.e., that issues are manually classified using only the title, but with different certainties. This data can only be generated by manual validation and is not available in any of the data sets we use. \cite{zolkeply2019classifying} could not be replicated because the authors do not state which 60 keywords they used in their approach.

\subsection{Performance metrics}

We take pattern from the literature \citep[e.g.][]{Antoniol2008, chawla2015automated, Terdchanakul2017, Pandey2018, qin2018classifying, kallis2019ticket} and base our evaluation on the \RECALL, \PRECISION, and \FSCORE, which are defined as
\begin{equation*}
\begin{split}
\RECALL &= \frac{tp}{tp+fn} \\
\PRECISION &= \frac{tp}{tp+fp} \\
\FSCORE &= 2 \cdot \frac{recall \cdot
precision}{recall+precision}
\end{split}
\end{equation*}
where $tp$ are true positives, i.e., bugs that are classified as bugs, $tn$ true negatives, i.e., non bugs classified as non bugs, $fp$ bugs not classified as bugs and $fn$ non bugs classified as bugs. The \RECALL{} measures the percentage of bugs that are correctly identified as bugs. Thus, a high \RECALL{} means that the model correctly finds most bugs. The \PRECISION{} measures the percentage of bugs among all predictions of bugs. Thus, a high \PRECISION{} means that there is strong likelihood that issues that are predicted as bugs are actually bugs. The \FSCORE{} is the harmonic mean of \RECALL{} and \PRECISION. Thus, a high \FSCORE{} means that the model is good at both predicting all bugs correctly and at not polluting the predicted bugs with two many other issues.

\subsection{Methodology}

\begin{figure}
\centering
\includegraphics[width=\textwidth]{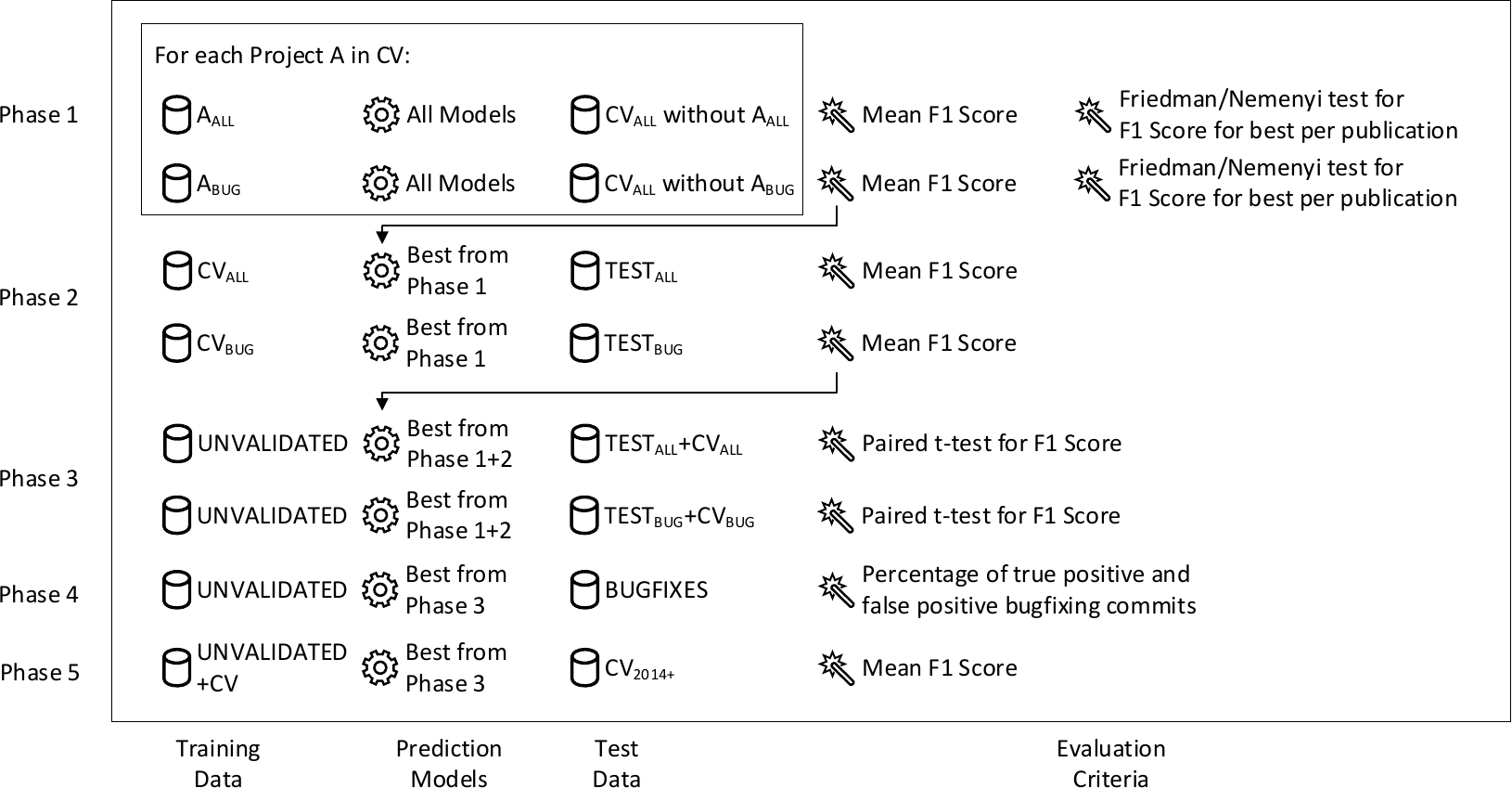}
\caption{Overview of the experiment methodology. The training and evaluation in all phases is conducted with all issues and with only bug issues. }
\label{fig:experiment-design}
\end{figure}

Figure~\ref{fig:experiment-design} summarizes our general methodology for the experiments, which consists of four phases. In Phase 1, we conduct a leave-one-project-out cross validation experiment with the CV data. This means that we use each project once as test data and train with all other projects. We determine the \RECALL, \PRECISION, and \FSCORE{} for all ten models we propose in Section~\ref{sec:putting-it-together} as well as all baselines this way for both the CV$_{ALL}$ and the CV$_{BUG}$ data. In case there are multiple variants, e.g., our ten approaches or different classifiers proposed for a baseline, we select the one that has the best overall mean value on the CV$_{ALL}$ and the CV$_{BUG}$ data combined. This way, we get a single model for each baseline, as well as for our approach, that we determined works best on the CV data. We then follow the guidelines from \cite{Demsar2006} for the comparison of multiple classifiers. Since the data is almost always normal, except for trivial models that almost always yield 0 as performance value, we report the mean value, standard deviation, and the confidence interval of the mean value of the results. The confidence interval with a confidence level of $\alpha$ for normally distributed samples is calculated as
\begin{equation}
mean \pm \frac{sd}{\sqrt{n}}Z_\alpha
\end{equation}
where the $sd$ is the standard deviation, $n$ the sample size, and $Z_\alpha$ the $\frac{alpha}{2}$ percentile of the t-distribution with $n-1$ degrees of freedom. However, the variances are not equal, i.e., the assumption of homoscedacity is not fulfilled. Therefore, we use the Friedman test~\citep{Friedman1940} with the post-hoc Nemenyi test~\citep{Nemenyi1963} to evaluate significant differences between the issue prediction approaches. The Friedman test is an omnibus test that determines if there is any difference in the central tendency of a group of paired samples with equal group sizes. If the outcome of the Friedman test is significant, the Nemenyi test evaluates which differences between approaches are significant based on the critical distance, which is defined as
\begin{equation}
    CD = \sqrt{\frac{k(k+1)}{12N}} q_{\alpha, N} 
\end{equation}
where $k$ is the number of approaches that are compared, $N$ is the number of distinct values for each approach, i.e., in our case the number of projects in a data set, and $q_{\alpha, N}$ is the $\alpha$ percentile of the studentized range distribution for $N$ groups and infinite degrees of freedom. Two approaches are significantly different, if the difference in the mean ranking between the performance of the approaches is greater than the critical distance. We use Cohen's $d$~\citep{Cohen1988} which is defined as
\begin{equation}
d = \frac{mean_1-mean_2}{\sqrt{\frac{sd_1+sd_2}{2}}}
\end{equation}
to report the effect sizes in comparison to the best performing approach. Table~\ref{tbl:effect_sizes} shows the magnitude of the effect sizes for Cohen's $d$. 
We will use the results from the first phase to evaluate RQ1, i.e., to see if our rules improved the issue type prediction. Moreover, the performance values will be used as indicators for RQ3.

\begin{table}[]
\centering
\begin{tabular}{ll}
\toprule
$d$    & Magnitude \\
\midrule
$d<0.2$ & Negligible \\
$0.2\leq d<0.5$ & Small \\
$0.5\leq d<0.8$ & Medium \\
$0.8\leq d$ & Large \\
\bottomrule
\end{tabular}
\caption{Magnitude of effect sizes of Cohen's $d$.}
\label{tbl:effect_sizes}
\end{table}

In Phase 2, we use the CV data as training to train a single model for the best performing approaches from Phase 1. This classifier is then applied to the TEST data. We report the mean value and standard deviation of the results. However, we do not conduct any statistical tests, because there are only five projects in the TEST data, which is insufficient for a statistical analysis. Through the results of Phase 2 we will try to confirm if the results from Phase 1 hold on unseen data. Moreover, we get insights into the reliability of the manually validated data, since different teams of researchers validated the CV and the TEST data. In case the performance is stable, we have a good indication that our estimated performance from Phase 1 generalizes. This is especially important, because Phase 2 is biased in favor of the state of the art, while Phase 1 is biased in favor of our approach. The reason for this is that most approaches from the state of the art were developed and tuned on the TEST data, while our approach was developed and tuned on the CV data. Therefore, the evaluation on the TEST data also serves as counter evaluation to ensure that the results are not biased due to the data that was used for the development and tuning of an approach, as stable results across the data sets would indicate that this is not the case. Thus, the results from Phase 2 are used to further evaluate RQ3 and to increase the validity of our results. 

In Phase 3, we evaluate the use of unvalidated data for the training, i.e., data about issues where the labels were not manually validated by researchers. For this, we compare the results of the best approach from Phase 1 and Phase 2 with the same approach, but trained with the UNVALIDATED data. Through this, we analyze RQ2 to see if we really require validated data or if unvalidated data works as well. Because the data is normal, we use the paired t-test to test if the differences between results in the \FSCORE{} are significant and Cohen's $d$ to calculate the effect size. Moreover, we consider how the \RECALL{} and \PRECISION{} are affected by the unvalidated data, to better understand how training with the UNVALIDATED data affects the results. 

We apply the approach we deem best suited in Phase 4 to a relevant problem of the Scenario 2 for issue type prediction discussed in Section~\ref{sec:terminology}. Concretely, we analyze if issue type prediction can be used to improve the identification of bugfixing commits. \cite{Herbold2020} found that mislabelled issues are the main source for the wrong identification of bugfixing commits. Therefore, an accurate issue type prediction model would be very helpful for any software repository mining tasks that relies on bugfixing commits. To evaluate the impact of the issue type prediction on the identification of bug fixing commits, we use two metrics. First, the percentage of actual bug fixing commits, that are found if issue type prediction is used (true positives). This is basically the same as the \RECALL{} of bugfixing commits. Second, the percentage of additional bugfixing commits that are found in additionally in relation to the actual number of bug fixing commits. This is indirectly related to the \PRECISION, because such additional commits are the result of false positive prediction. 

Finally, we apply the best approach in a setting that could be Scenario 3 or the Scenario 4 outlined in Section~\ref{sec:terminology}. An important aspect we ignored so far is the potential information leakage because we did not consider the time when issues were reported. In a realistic scenario where we apply the prediction live within an issue tracking system, data from the future is not available. Instead, only past issues may be used to train the model. For this scenario, we decide on a fixed cutoff date for the training data. All data prior to this cutoff is used for training, all data afterwards for testing of the prediction performance. This is realistic, because such models are often trained at some point and the trained model is then implemented in the live system and must be actively updated as a maintenance task. We use the UNVALIDATED+CV data for training and the CV$_{2014+}$ for testing in this phase. We compare the results with the performance we measured in Phase 3 of the experiments, to understand how this realistic setting affects our performance estimations in comparison to the less realistic results that ignore the potential information leakage because of overlaps in time between the training and test data. 

For our experiments, we conduct many statistical tests. We use Bonferroni correction~\citep{Dunn1961} to account for false positive due to the repeated tests and have an overall significance level of 0.05, resp. confidence level of 0.95 for the confidence intervals. We use a significance level of $\frac{0.05}{22} = 0.0023$ for the Shapiro-Wilk tests for normality~\citep{Shapiro1965}, because we perform nine tests for normality for the best performing approaches for both all issues and only bugs in both Phase 1 and four additional tests for normality in Phase 3. We conduct two Bartlett tests for homoscedacity~\citep{Bartlett1937} in Phase 1 with a significance level of $\frac{0.05}{2} = 0.025$. We conduct four tests for the significance of differences between classifiers with a significance level of $\frac{0.05}{4}$, i.e., two Friedman tests in Phase 1 and two paired t-tests in Phase 3. Moreover, we calculate the confidence intervals for all results in Phase 1 and Phase 3, i.e., 25 results for both CV$_{ALL}$ and CV$_{BUG}$ and four results for Phase 3. Hence, we use a confidence level of $1-\frac{0.05}{54} = 0.999$ for these confidence intervals. 

\subsection{Results}
\label{sec:results}

\subsubsection{Results for Phase 1}

\begin{figure*}
\centering
\begin{tabular}{>{\rowmac}l>{\rowmac}r>{\rowmac}r>{\rowmac}r>{\rowmac}r<{\clearrow}}
\toprule
{} &  mean &    sd &              CI &  $d$ \\
\midrule \setrow{\bfseries}
Herbold2020-FTA        & 0.643 & 0.077 &  [0.598, 0.689] &   0.000 \\
Herbold2020-FTA+NPE      & 0.639 & 0.073 &  [0.596, 0.681] &   0.064 \\
Herbold2020-FT         & 0.627 & 0.072 &  [0.585, 0.669] &   0.218 \\
Herbold2020-FT+NPE       & 0.619 & 0.072 &  [0.577, 0.661] &   0.326 \\
Herbold2020-RF         & 0.313 & 0.092 &  [0.260, 0.367] &   3.896 \\
Herbold2020-RF+NPE       & 0.307 & 0.077 &  [0.262, 0.352] &   4.356 \\
Herbold2020-Basic-RF   & 0.222 & 0.104 &  [0.161, 0.283] &   4.606 \\
Herbold2020-NB+NPE       & 0.169 & 0.060 &  [0.134, 0.204] &   6.861 \\
Herbold2020-Basic-NB   & 0.017 & 0.029 &  [0.000, 0.033] &  10.759 \\
Herbold2020-NB & 0.015 & 0.018 &  [0.005, 0.026] &  11.205 \\
\midrule \setrow{\bfseries}
Kallis2019-FT          & 0.638 & 0.074 &  [0.594, 0.681] &   0.073 \\
\midrule \setrow{\bfseries}
Pandey2018-LR          & 0.541 & 0.070 &  [0.500, 0.582] &   1.387 \\
Pandey2018-NB          & 0.503 & 0.111 &  [0.439, 0.568] &   1.469 \\
\midrule
Qin2018-LSTM           & 0.349 & 0.096 &  [0.293, 0.405] &   3.370 \\
\midrule \setrow{\bfseries}
Trivial                & 0.349 & 0.096 &  [0.293, 0.405] &   3.370 \\
\midrule \setrow{\bfseries}
Chawla2015-FL          & 0.208 & 0.070 &  [0.168, 0.249] &   5.913 \\
\midrule \setrow{\bfseries}
Palacio2019-SRN         & 0.206 & 0.068 &  [0.166, 0.246] &   6.017 \\
\midrule \setrow{\bfseries}
Limsettho2014-LR       & 0.154 & 0.100 &  [0.096, 0.212] &   5.490 \\
Limsettho2014-NB       & 0.000 & 0.001 &  [0.000, 0.001] &  11.777 \\
\midrule \setrow{\bfseries}
Terdchanakul2017-RF    & 0.248 & 0.088 &  [0.196, 0.299] &   4.775 \\
Terdchanakul2017-LR    & 0.076 & 0.068 &  [0.037, 0.116] &   7.796 \\
\midrule \setrow{\bfseries}
Otoom2019-NB           & 0.059 & 0.077 &  [0.013, 0.104] &   7.561 \\
Otoom2019-RF           & 0.010 & 0.024 &  [0.000, 0.024] &  11.070 \\
Otoom2019-SVC          & 0.003 & 0.010 &  [0.000, 0.008] &  11.638 \\
\midrule \setrow{\bfseries}
Pingclasai2013-LR      & 0.003 & 0.009 &  [0.000, 0.009] &  11.636 \\
Pingclasai2013-NB      & 0.004 & 0.011 &  [0.000, 0.010] &  11.599 \\
\bottomrule
\end{tabular}
%\end{minipage}
\hspace{0.5cm}
\begin{minipage}{\textwidth}
\includegraphics[width=0.5\textwidth]{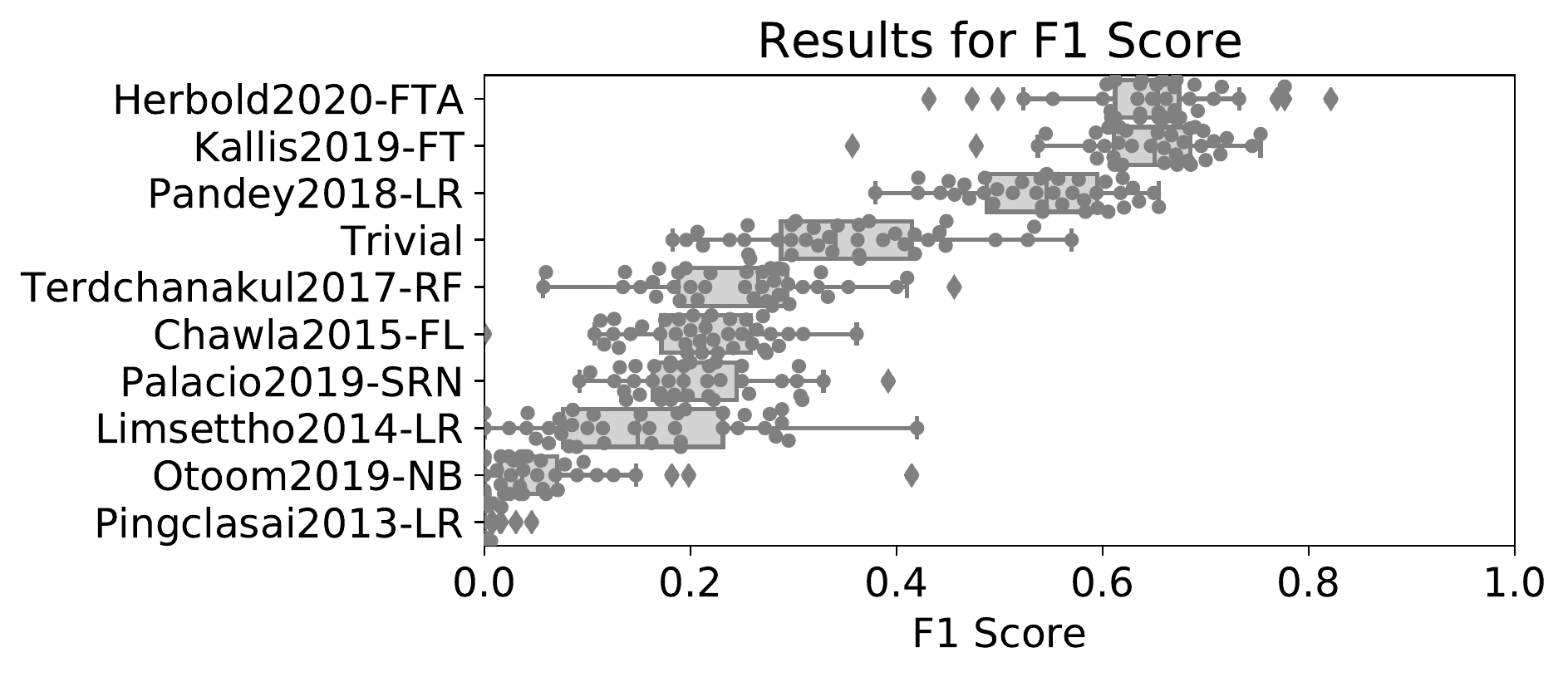}
\includegraphics[width=0.5\textwidth]{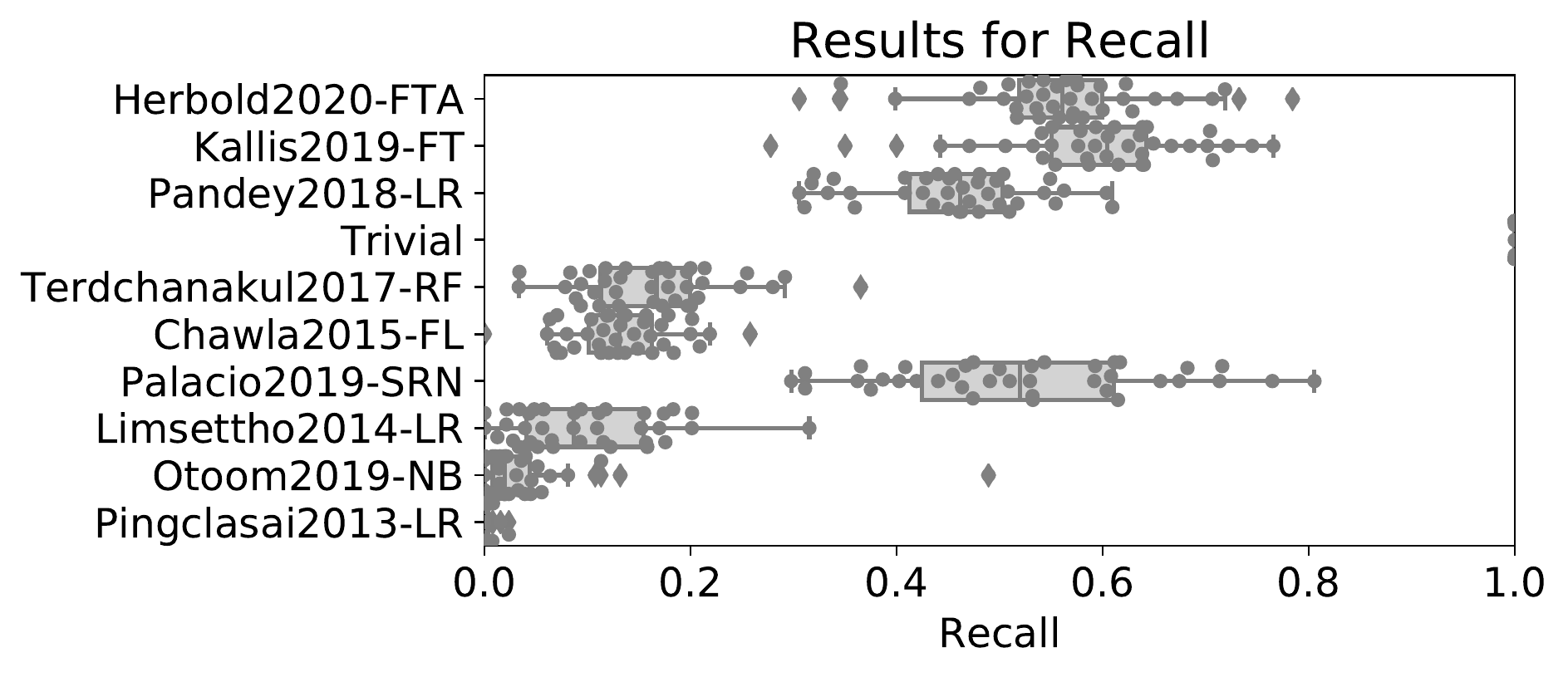}
\includegraphics[width=0.5\textwidth]{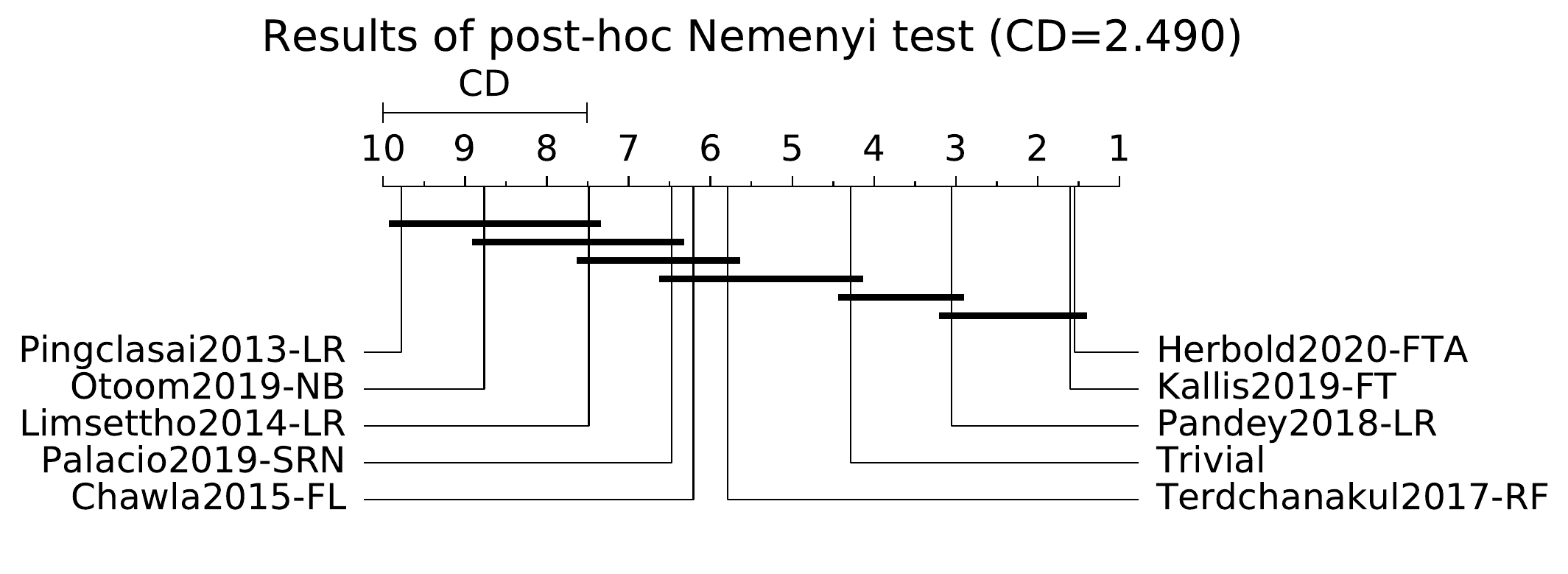}
\includegraphics[width=0.5\textwidth]{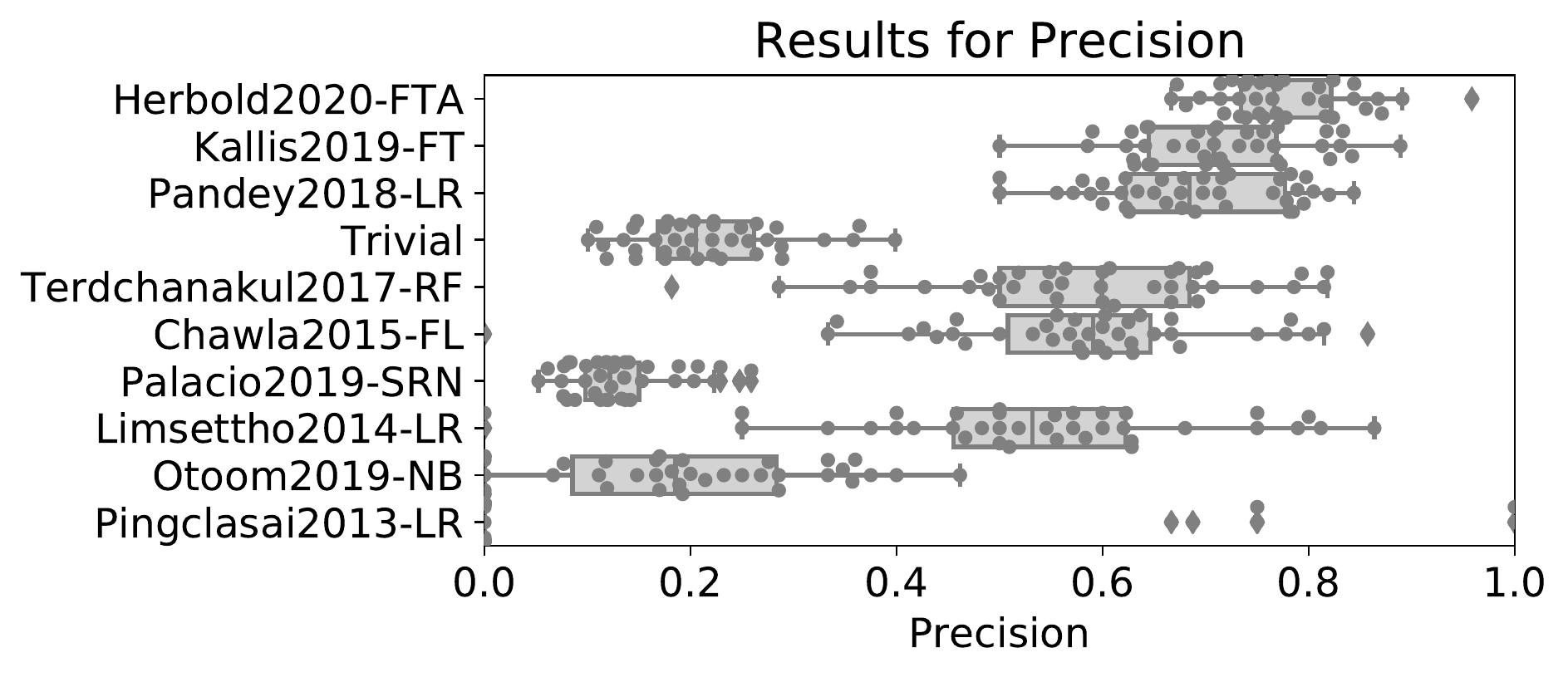}
\end{minipage}
\caption{Results of leave-one-project-out cross validation with the CV$_{ALL}$ data. The bold-faced approaches where the best for a publication and are used in the statistical analysis.}
\label{fig:phase1_all_issues}
\end{figure*}

\begin{figure*}
\centering
\begin{tabular}{>{\rowmac}l>{\rowmac}r>{\rowmac}r>{\rowmac}r>{\rowmac}r<{\clearrow}}
\toprule
{} &  mean &    sd &              CI &  $d$ \\
\midrule
Herbold2020-RF         & 0.809 & 0.051 &  [0.779, 0.839] &  -0.082 \\
\setrow{\bfseries}
Herbold2020-FTA        & 0.805 & 0.049 &  [0.776, 0.834] &   0.000 \\
Herbold2020-FT         & 0.803 & 0.050 &  [0.774, 0.832] &   0.041 \\
Herbold2020-RF+NPE       & 0.803 & 0.046 &  [0.776, 0.830] &   0.044 \\
Herbold2020-FT+NPE       & 0.801 & 0.046 &  [0.774, 0.828] &   0.088 \\
Herbold2020-FTA+NPE      & 0.802 & 0.048 &  [0.774, 0.830] &   0.057 \\
Herbold2020-NB+NPE       & 0.799 & 0.067 &  [0.760, 0.839] &   0.098 \\
Herbold2020-NB         & 0.790 & 0.066 &  [0.751, 0.829] &   0.252 \\
Herbold2020-Basic-NB   & 0.767 & 0.083 &  [0.718, 0.816] &   0.563 \\
Herbold2020-Basic-RF   & 0.776 & 0.056 &  [0.743, 0.809] &   0.546 \\
\midrule \setrow{\bfseries}
Kallis2019-FT          & 0.777 & 0.062 &  [0.740, 0.814] &   0.503 \\
\midrule \setrow{\bfseries}
Pandey2018-LR          & 0.760 & 0.058 &  [0.726, 0.794] &   0.839 \\
Pandey2018-NB          & 0.765 & 0.069 &  [0.725, 0.806] &   0.667 \\
\midrule \setrow{\bfseries}
Pandey2018-NB          & 0.765 & 0.069 &  [0.725, 0.806] &   0.667 \\
\midrule
Qin2018-LSTM           & 0.729 & 0.083 &  [0.681, 0.778] &   1.115 \\
\midrule \setrow{\bfseries}
Trivial                & 0.729 & 0.083 &  [0.681, 0.778] &   1.115 \\
\midrule \setrow{\bfseries}
Otoom2019-SVC          & 0.722 & 0.074 &  [0.678, 0.765] &   1.324 \\
Otoom2019-NB           & 0.718 & 0.075 &  [0.674, 0.762] &   1.382 \\
Otoom2019-RF           & 0.718 & 0.071 &  [0.676, 0.760] &   1.429 \\
\midrule \setrow{\bfseries}
Pingclasai2013-LR      & 0.715 & 0.085 &  [0.665, 0.765] &   1.297 \\
Pingclasai2013-NB      & 0.671 & 0.131 &  [0.594, 0.748] &   1.361 \\
\midrule \setrow{\bfseries}
Limsettho2014-LR       & 0.670 & 0.103 &  [0.610, 0.731] &   1.671 \\
Limsettho2014-NB       & 0.734 & 0.078 &  [0.688, 0.780] &   1.090 \\
\midrule 
Terdchanakul2017-LR    & 0.728 & 0.082 &  [0.680, 0.777] &   1.136 \\
\setrow{\bfseries}
Terdchanakul2017-RF    & 0.610 & 0.103 &  [0.550, 0.671] &   2.425 \\
\midrule \setrow{\bfseries}
Palacio2019-SRN         & 0.290 & 0.121 &  [0.218, 0.361] &   5.600 \\
\bottomrule
\end{tabular}
\hspace{0.5cm}
\begin{minipage}{\textwidth}
\includegraphics[width=0.5\textwidth]{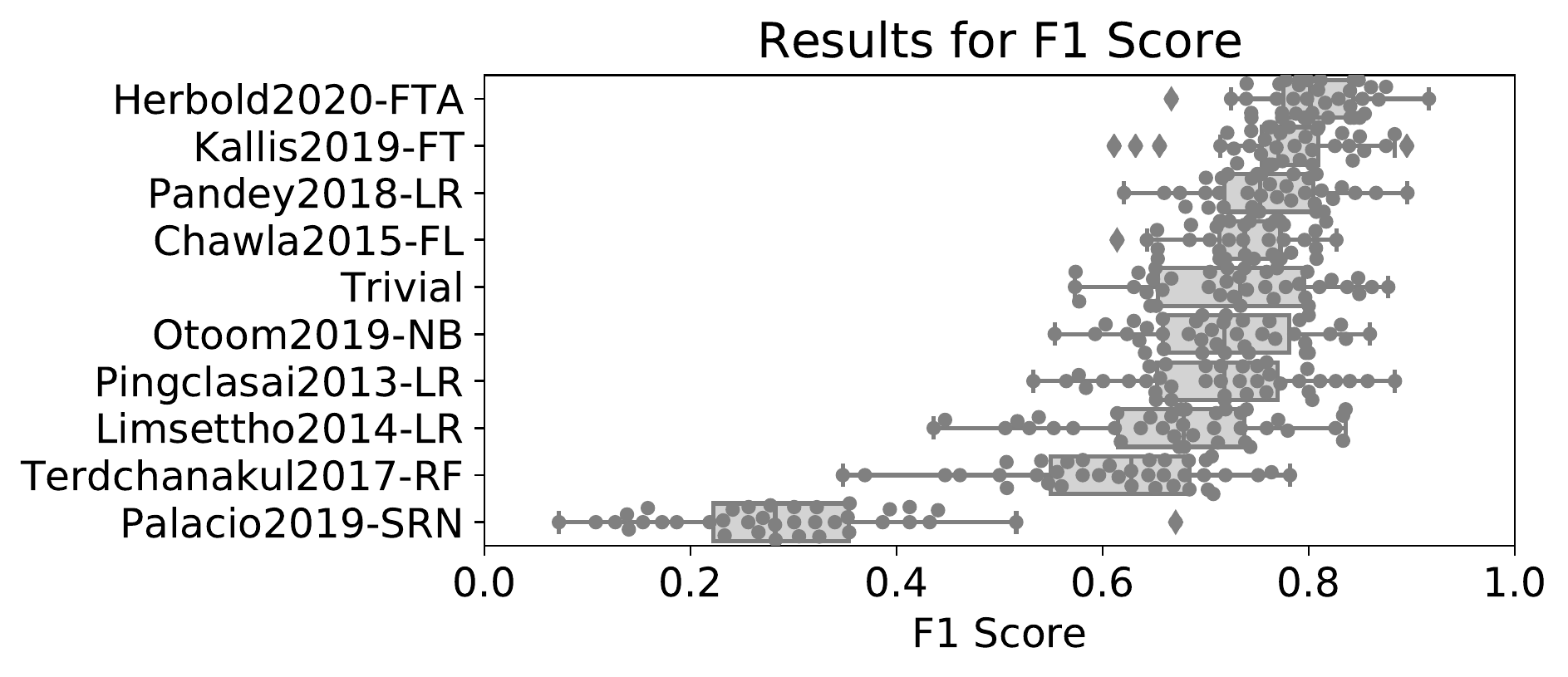}
\includegraphics[width=0.5\textwidth]{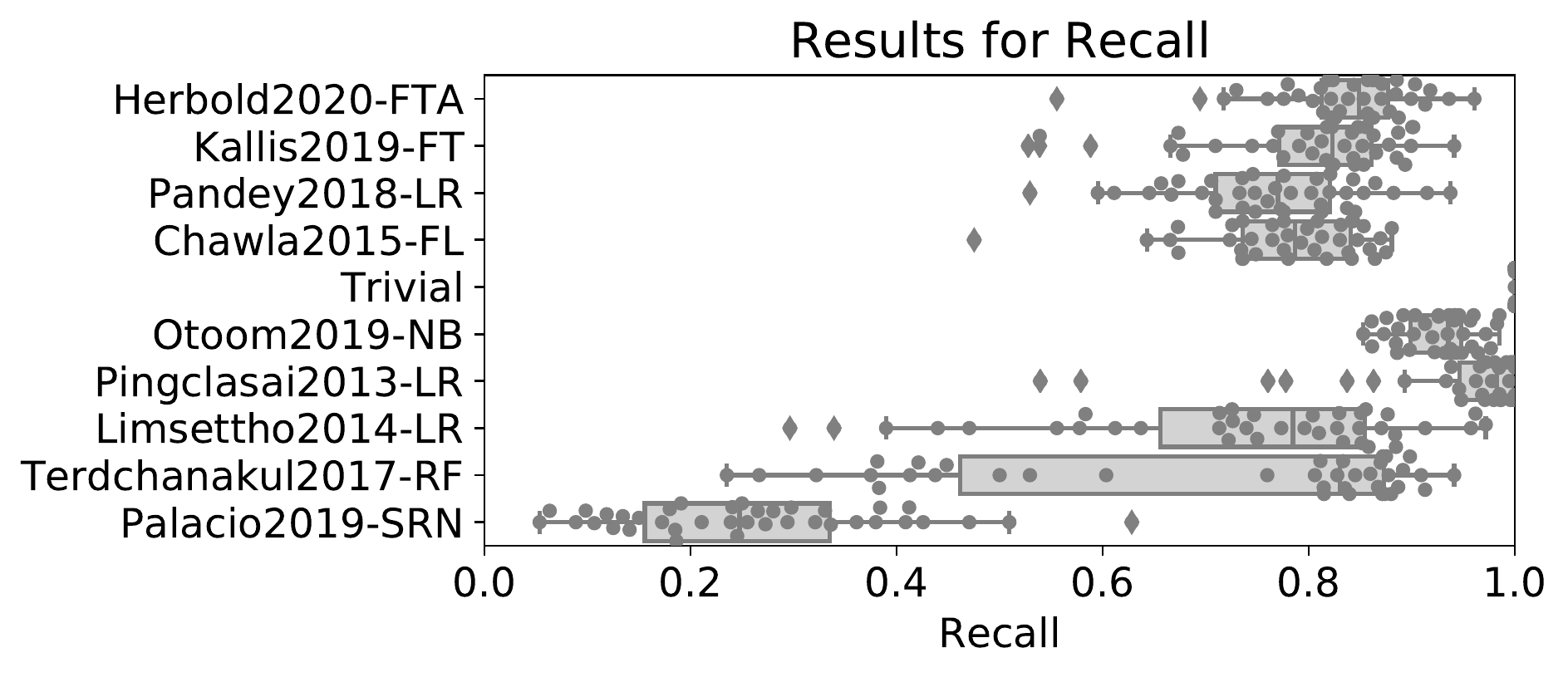}
\includegraphics[width=0.5\textwidth]{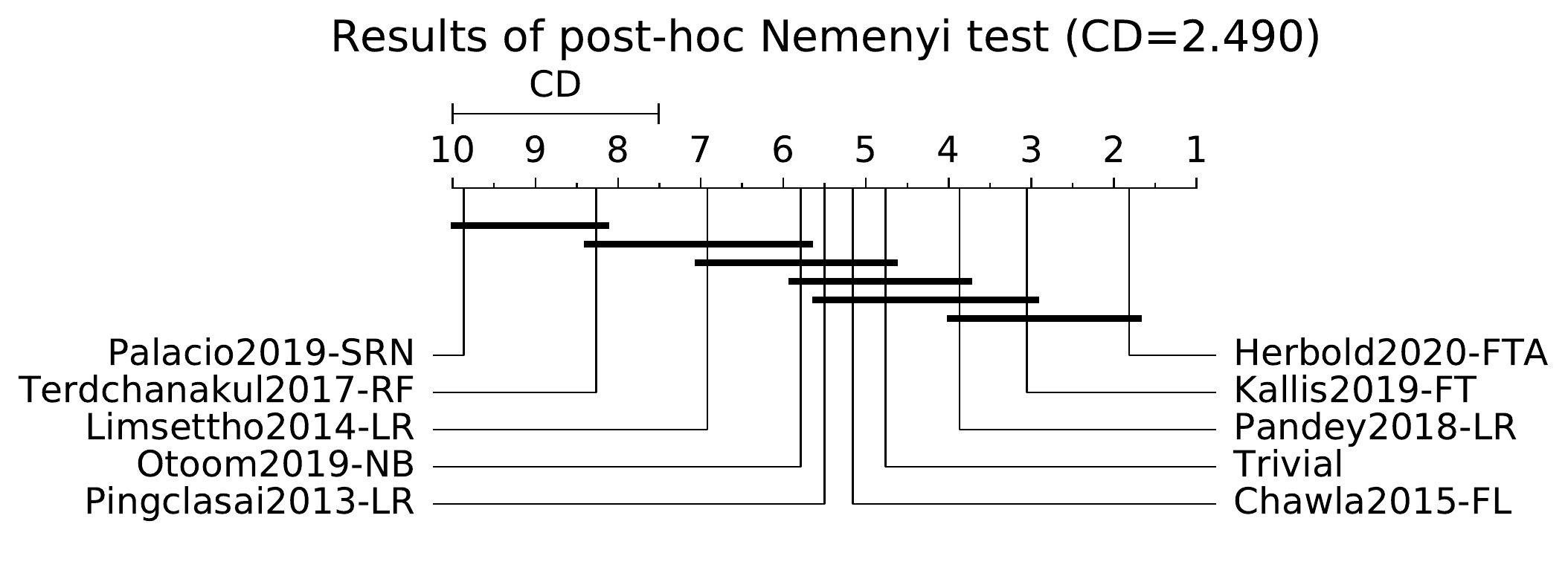}
\includegraphics[width=0.5\textwidth]{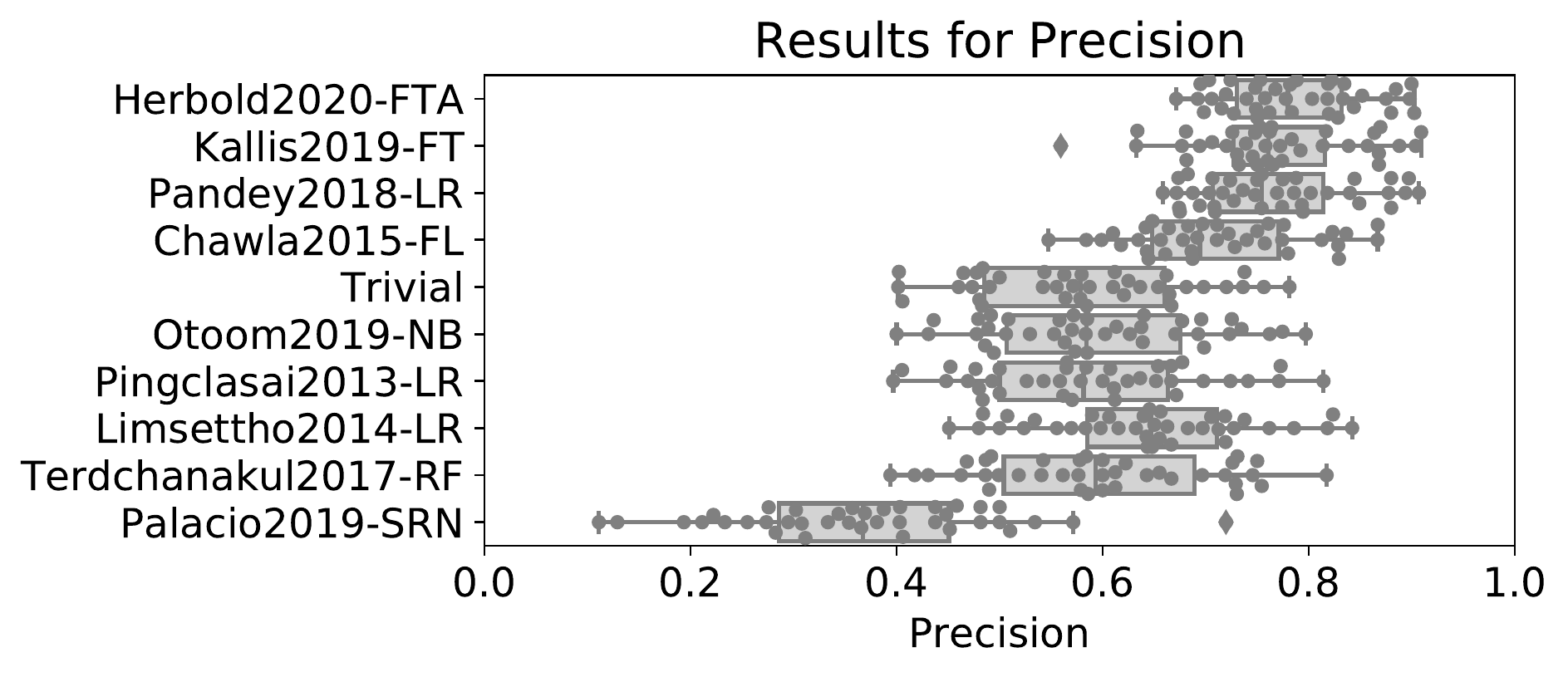}
\end{minipage}
\caption{Results of leave-one-project-out cross validation with the CV$_{BUG}$ data. The bold-faced approaches where the best for a publication and are used in the statistical analysis.}
\label{fig:phase1_bugs_only}
\end{figure*}

Figures~\ref{fig:phase1_all_issues} and~\ref{fig:phase1_bugs_only} show the the results for the first phase of the experiment on the CV$_{ALL}$ and CV$_{BUGS}$ data, respectively. We observe that while there is a strong variance in the \FSCORE{} using the CV$_{ALL}$ data with values between 0.0 (Limsettho2014-NB) and 0.643 (Herbold2020-FTA), the results on the CV$_{BUG}$ data are more stable with values between 0.610 (Terdchanakul2017-RF) and 0.809 (Herbold2020-RF). The strong performance on CV$_{BUG}$ includes the Trival approach, i.e., simply predicted everything as bug is already relatively hard to beat, because the \RECALL{} is perfect and the \PRECISION{} is at about 60\%.\footnote{60\% is roughly the amount of bugs within the data} This finding is different for the CV$_{ALL}$, because here the class level imbalance is the other way around and the \PRECISION{} drops to roughly 20\%. In general, we observe that the \FSCORE{} with the CV$_{ALL}$ data is lower than with CV$_{BUG}$. 

This also shows in the results for the different variants we proposed in Section~\ref{sec:putting-it-together}, were we observe big difference with the CV$_{ALL}$ data and only relatively small differences with the CV$_{BUG}$ data. On the CV$_{ALL}$ data, the strongest driver of the differences is the choice of the classifier. The models using FTA perform best, followed by FT classifier which has a slightly worse performance. The drop between FT and RF is steep, NB performs worse and predicts only few bugs. Using distinct classifiers for the title and description improves the performance slightly. However, additional classifiers for null pointers lead to slightly worse results. Thus, we find that Herbold2020-FTA with separate classifiers for title and descriptions performs best for CV$_{ALL}$, even though the difference to the other variants with FTA/FT is small. On the CV$_{BUG}$, the \FSCORE{} of all models that use different separate classifiers for title and description is within the interval $[0.790, 0.809]$. 

\begin{figure}
\begin{minipage}{0.49\textwidth}
\centering
{\scriptsize Results for $CV_{ALL}$}\\
\includegraphics[width=\textwidth]{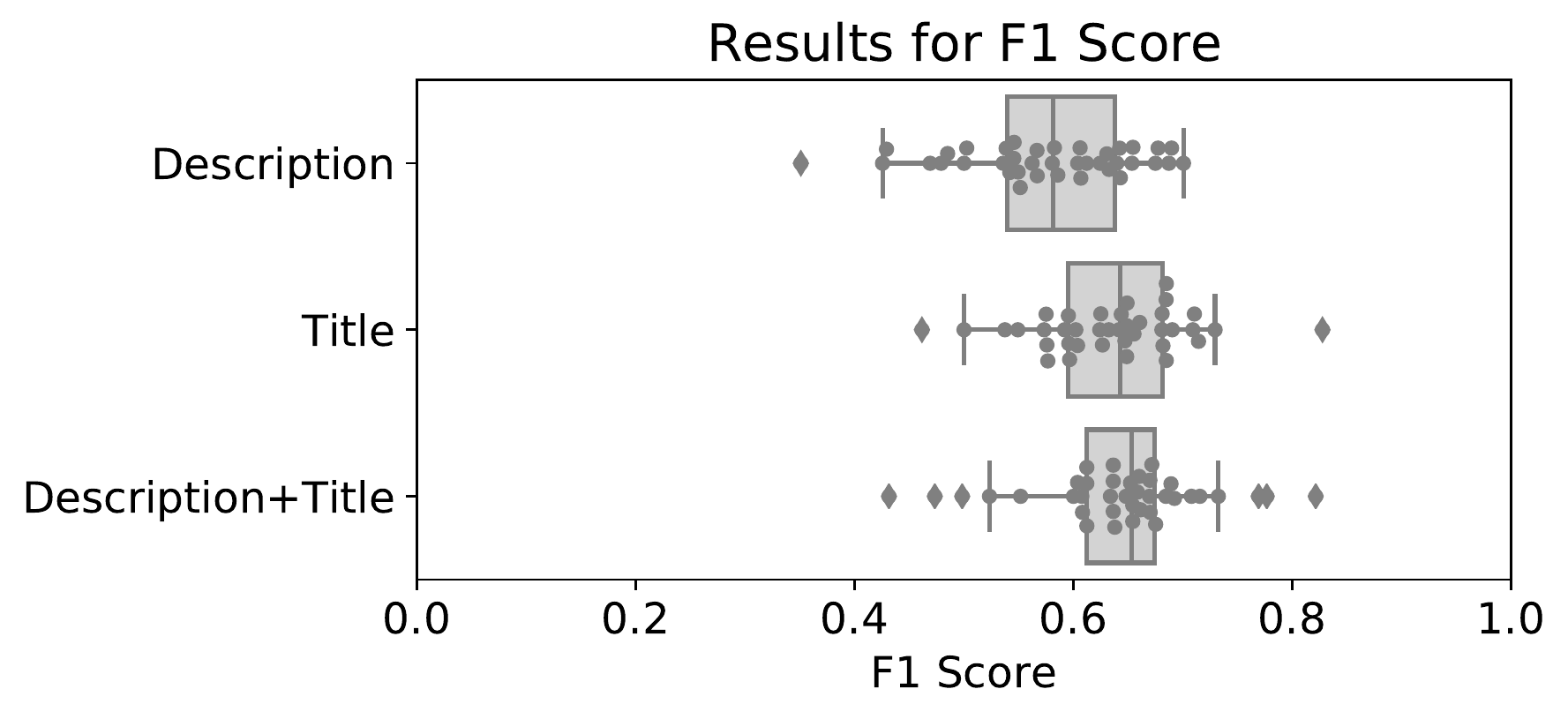}
\end{minipage}
\begin{minipage}{0.49\textwidth}
\centering
{\scriptsize Results for $CV_{BUG}$}\\
\includegraphics[width=\textwidth]{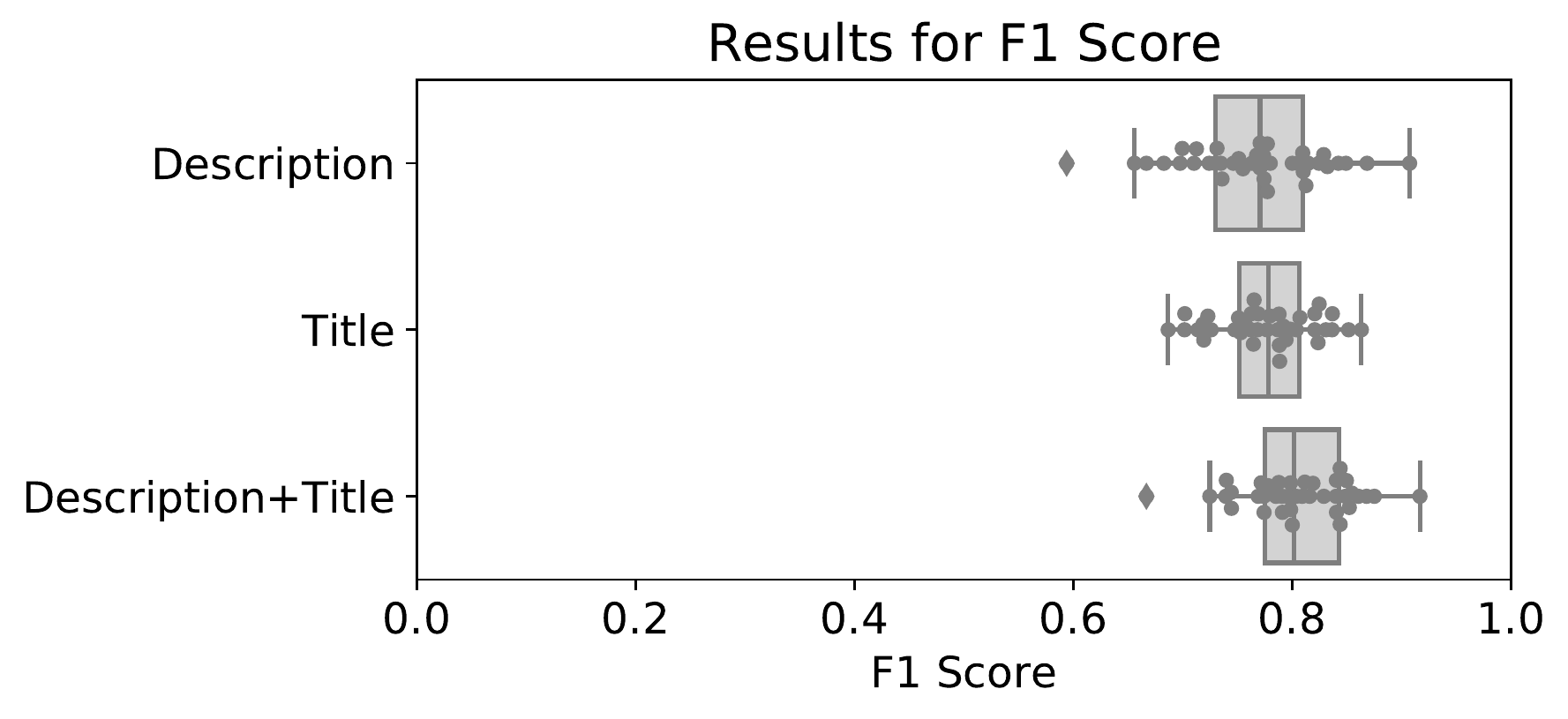}
\end{minipage}
\caption{Results for the different classifiers for title and description in comparison to their combination to gain insights into the Herbold2020-FTA model.}
\label{fig:insights}
\end{figure}

Thus, we find that overall, Herbold2020-FTA performs best among the approaches discussed in Section~\ref{sec:putting-it-together}. Figure~\ref{fig:insights} allows us to gain further insights and to understand how the approach achieves the performance. The figure shows the performance of the two predictors that are internally used in comparison to the overall performance. We see that for both the $CV_{ALL}$ and the $CV_{BUG}$ data, that the performance of using only the title or description yields worse results than the combination of both classifiers. This indicates that the differences between title and description we describe in Section~\ref{sec:title-desc} are indeed relevant. Moreover, because the combination of both classifiers outperforms each classifier on its own, this indicates that some issues that are hard to predict based on title can be predicted based on the description, and vice versa.

For the related work that suggested multiple classifiers, our observations are similar. The differences on the CV$_{BUG}$ data are small, the performance on the CV$_{ALL}$ data is always worse and, in many cases, very bad with only very few bugs predicted. The only outlier is Palacio2019-SRN, which performs better on the CV$_{ALL}$ data than on the CV$_{BUG}$, but relatively bad on both data sets. We hypothesize that this is because there are fewer training issues available in CV$_{BUG}$ and see this as an indicator that Palacio2019-SRN may perform better in future benchmarks, given that the amount of data is increased.

Of the approaches from the related work, only Kallis2019-FT and Pandey2018-LR achieve an \FSCORE{} of over 0.5 on CV$_{ALL}$. We also note that the approach Qin2018-LSTM degenerated into a trivial model that predicts everything as bug. Consequently, we exclude Qin2018-LSTM from our subsequent statistical analysis, because we would, otherwise, have the same model twice. 

We reject the null hypothesis of the Friedman test that there are no significant differences between the approaches on the CV$_{ALL}$ data ($p-value<0.001$). The post-hoc Nemenyi test found that while our Herbold2020-FTA approach has the best mean ranking, the difference is not significant in comparison to Kallis2019-FT and Pandey2018-LR. The difference in \FSCORE{} between our Herbold2020-FTA and Kallis2019-FT is almost non-existent. Thus, the advantage of automatically tuning the FT classifier and the usage of different classifiers for title and description is very small. The magnitude of the (non-significant) effect size between Herbold2020-FTA and Kallis2019-FT is negligible with $d=0.073$. However, there is a difference between the \RECALL{} and \PRECISION{} of Herbold2020-FTA and Kallis2019-FT. Herbold2020-FTA has a higher \PRECISION, while Kallis2019-FT has a higher \RECALL. While not significant, the difference between Pandey2018-LR on the one hand and Herbold2020-FTA and Kallis2019-FT on the other hand is larger, both in the mean value and also in the (non-significant) effect size which has a large magnitude with $d=1.387$. All other models from the state of the art are significantly different from Herbold2020-FTA and Kallis2019-FT with a large effect size of at least $d=3.370$. We note that Pandey2018-LR is not significantly different from the trivial model. Moreover, we observe that all other approaches are all equal to or worse than trivially predicting everything as bug in the mean \FSCORE, because of a low \RECALL. Pinglasai2013-LR, Otoom2019-NB, and Limsettho2014-LR are even significantly worse than the trivial model. 

We reject the null hypothesis of the Friedman test that there are no significant differences between the approaches on the CV$_{BUG}$ data ($p-value<0.001$).  The post-hoc Nemenyi test found that our Herbold2020-FTA has the best best mean ranking, but the diffference is not significant in comparison to Kallis2019-FT and Pandey2018-LR. However, in comparison to the CV$_{ALL}$ data, there is some gap between Herbold2020-FTA and Kallis2019-FT both in the mean \FSCORE{} and the magnitude of the effect size would be medium with $d=0.503$ if it where significant. Moreover, Herbold2020-FTA yields a slightly better performance in both \RECALL{} and \PRECISION, i.e., there is no trade-off between Herbold2020-FTA and Kallis2019-FT between \RECALL{} and \PRECISION. The gap between Herbold2020-FTA and Panday2018-LR is similar to the gap on the CV$_{ALL}$ of effect and the effect size would be large with $d=0.839$. All other approaches from the state of the art are significantly worse than Herbold2020-FTA with a large effect size of at least $d=1.115$. We note that Kallis2019-FT and Pandey2018-LR are not significantly different from the trivial model. Same as for the CV$_{BUG}$ data, we find that all other approaches have a worse mean ranking than the trivial approach, even though Chawla2015-FL has a slightly higher mean value than the trivial model. However, only Terdchanakul2017-RF is significantly worse the the trivial model. 

When we consider the results on CV$_{BUG}$ and CV$_{ALL}$ together, the approaches Herbold2020-FTA, Kallis2019-FT and Pandey2018-LR are consistently ranked first, with their mean ranks in that order. The differences between these three approaches are not significant. However, we note that there is a considerable gap in the mean ranks reported on the CD diagrams between Herbold2020-FTA and Kallis2019-FT on the hand, and Pandey2018-LR on the other hand. This gap also shows in the mean \FSCORE, which is lower on both the CV$_{ALL}$ and CV$_{BUG}$ data. In fact, Pandey2018-LR almost always ranks worse on than both Herbold2020-FTA and Kallis2019-FT. Thus, we believe that the reason that Pandey2018-LR is not significantly different from Herbold2020-FTA and Kallis2019-FT is our limited sample size of 38 projects, which leads to a critical distance of 2.216 for the Nemenyi test. Thus, even if Herbold2020-FTA would always have the best score, Kallis2019-FT would always have the second best score, and Pandey2018-LR would always have the third best score, the difference would still not be significant. The reason for this is that we have not enough data, given that we are not just comparing these three populations, but a total of nine approaches at the same time. Therefore, we predict that with more data Herbold2020-FTA and Kallis2019-FT would significantly outperform Pandey2018-LR, but do not have the data to substantiate our claim. 

Given this prediction, we conclude from Phase 1 that Facebook AI Research did a very good job on the design of FT that outperforms other classification algorithms, but that the automated tuning may not always yield better results, e.g., because there is not enough data. Similarly, different classifiers for title and description may improve the results of FT, but not significantly. 

We also note that approaches that use NB as classifier perform a lot worse on the CV$_{ALL}$ data than on the CV$_{BUG}$ data. We believe this may be because the models are not actually learning a good scoring function, but rather relatively randomly guessing the number of bugs based on the a-priori probability of the class in the training data. This should work reasonably well in the CV$_{BUG}$ data because of the class level imbalance in favor of bug issues, but should fail in case of the CV$_{ALL}$ data because the chance of correctly hitting bugs when roughly 20\% of the issues are randomly predicted is relatively low.

\subsubsection{Results of Phase 2}

\begin{figure*}
\centering
\begin{tabular}{lrrrrr|rr}
\toprule
{} &  http- &  jack- &  lucene- &     rhino &    tomcat &      mean &       std \\
& comp. & rabbit & solr & & & & \\
\midrule
Herbold2020-FTA     &                  0.692 &       0.732 &        0.634 &  0.705 &   0.641 & 0.681 & 0.038 \\
Kallis2019-FT       &                  0.706 &       0.707 &        0.620 &  0.719 &   0.633 & 0.677 & 0.042 \\
Pandey2018-LR       &                  0.630 &       0.610 &        0.574 &  0.582 &   0.594 & 0.598 & 0.020 \\
Qin2018-LSTM        &                  0.580 &       0.566 &        0.446 &  0.682 &   0.708 & 0.596 & 0.093 \\
Palacio2019-SRN     &                  0.385 &       0.366 &        0.340 &  0.406 &   0.520 & 0.404 & 0.062 \\
Terdchanakul2017-RF &                  0.203 &       0.285 &        0.252 &  0.246 &   0.321 & 0.261 & 0.040 \\
Chawla2015-FL       &                  0.276 &       0.238 &        0.266 &  0.179 &   0.186 & 0.229 & 0.040 \\
Limsettho2014-LR    &                  0.192 &       0.278 &        0.045 &  0.202 &   0.384 & 0.221 & 0.111 \\
Otoom2019-NB        &                  0.037 &       0.064 &        0.016 &  0.026 &   0.092 & 0.047 & 0.028 \\
Pingclasai2013-LR   &                  0.000 &       0.000 &        0.000 &  0.000 &   0.000 & 0.000 & 0.000 \\
\bottomrule
\end{tabular}
\hspace{0.5cm}
\begin{minipage}{0.49\textwidth}
\includegraphics[width=\textwidth]{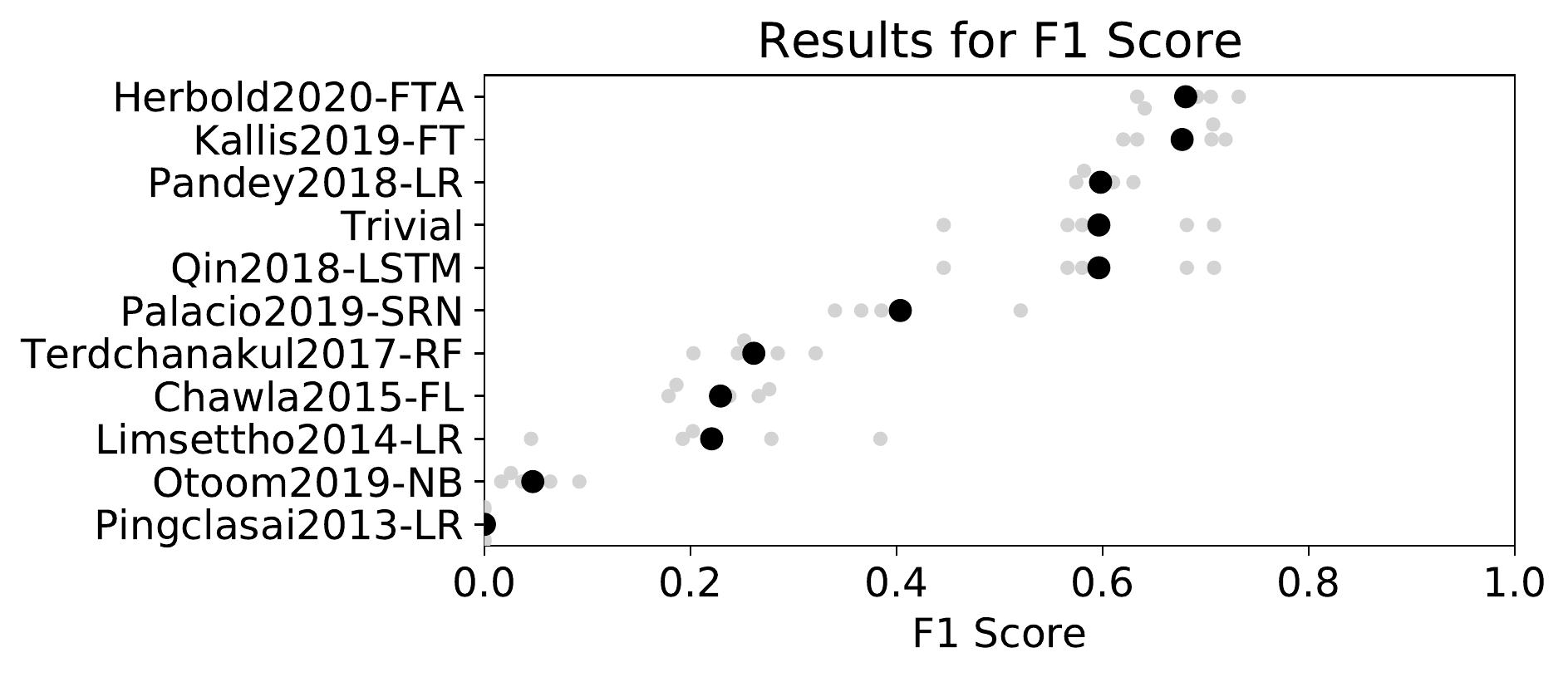}
\end{minipage}
\begin{minipage}{0.49\textwidth}
\includegraphics[width=\textwidth]{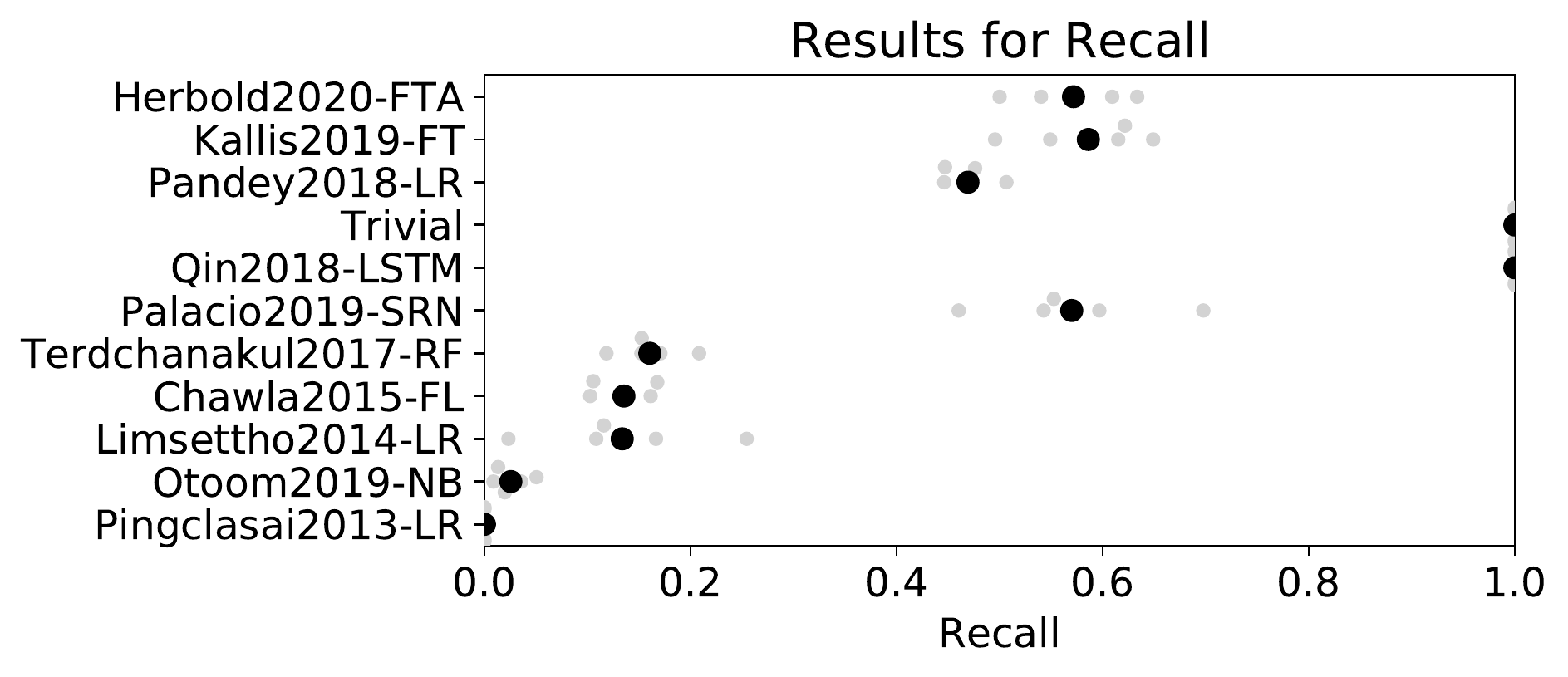}
\includegraphics[width=\textwidth]{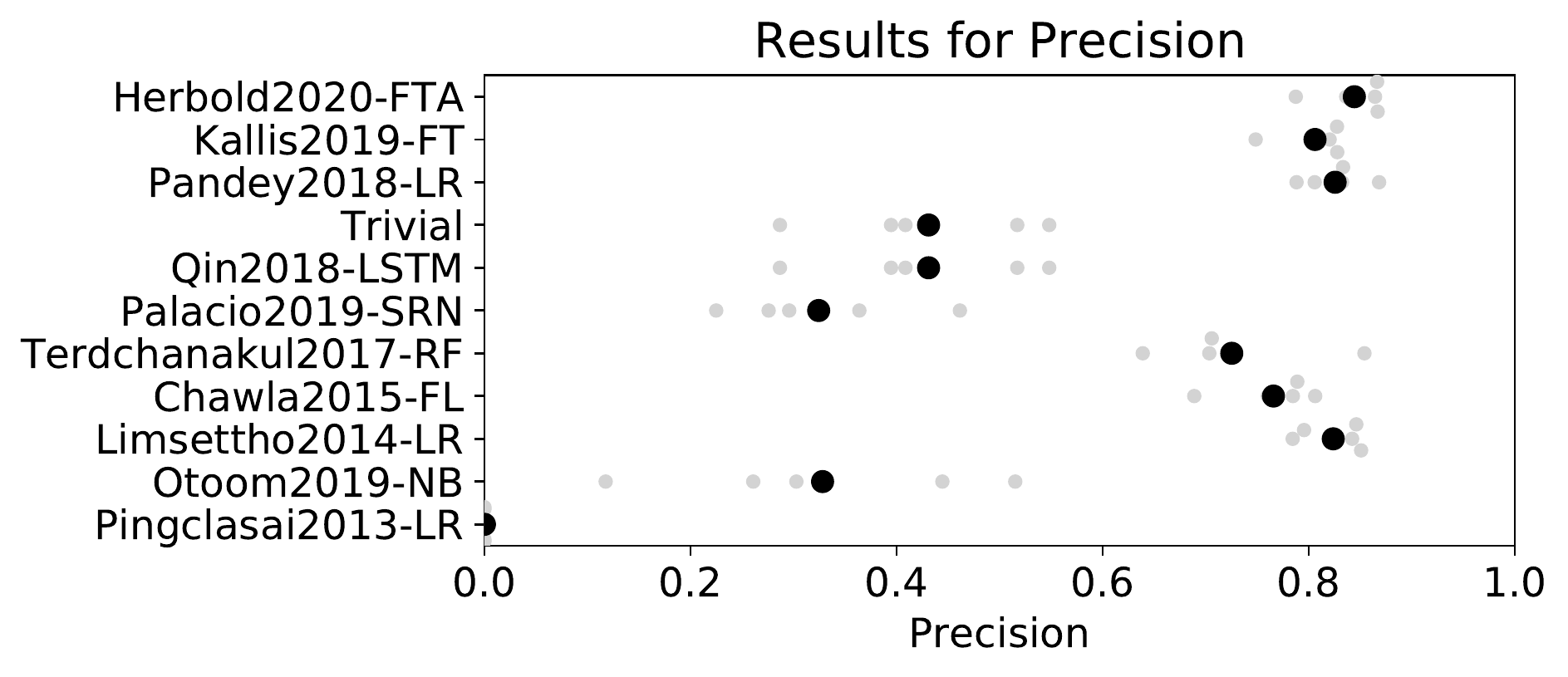}
\end{minipage}
\caption{Results of training with the CV$_{ALL}$ and testing with the TEST$_{ALL}$ data.}
\label{fig:phase2_all_issues}
\end{figure*}

\begin{figure*}
\centering
\begin{tabular}{lrrrrr|rr}
\toprule
{} &  http- &  jack- &  lucene- &     rhino &    tomcat &      mean &       std \\
& comp. & rabbit & solr & & & & \\
\midrule
Herbold2020-FTA     &                  0.832 &       0.874 &        0.777 &  0.813 &   0.783 & 0.816 & 0.035 \\
Kallis2019-FT       &                  0.821 &       0.841 &        0.767 &  0.803 &   0.783 & 0.803 & 0.026 \\
Qin2018-LSTM        &                  0.776 &       0.857 &        0.791 &  0.748 &   0.760 & 0.786 & 0.038 \\
Pandey2018-LR       &                  0.806 &       0.826 &        0.797 &  0.760 &   0.743 & 0.786 & 0.031 \\
Otoom2019-NB        &                  0.776 &       0.826 &        0.789 &  0.764 &   0.751 & 0.781 & 0.025 \\
Chawla2015-FL       &                  0.777 &       0.801 &        0.756 &  0.712 &   0.726 & 0.755 & 0.033 \\
Pingclasai2013-LR   &                  0.778 &       0.848 &        0.777 &  0.742 &   0.602 & 0.750 & 0.081 \\
Limsettho2014-LR    &                  0.765 &       0.774 &        0.691 &  0.732 &   0.759 & 0.744 & 0.030 \\
Terdchanakul2017-RF &                  0.561 &       0.546 &        0.442 &  0.690 &   0.717 & 0.591 & 0.101 \\
Palacio2019-SRN     &                  0.169 &       0.212 &        0.273 &  0.150 &   0.475 & 0.256 & 0.118 \\
\bottomrule
\end{tabular}
\hspace{0.5cm}
\begin{minipage}{0.49\textwidth}
\includegraphics[width=\textwidth]{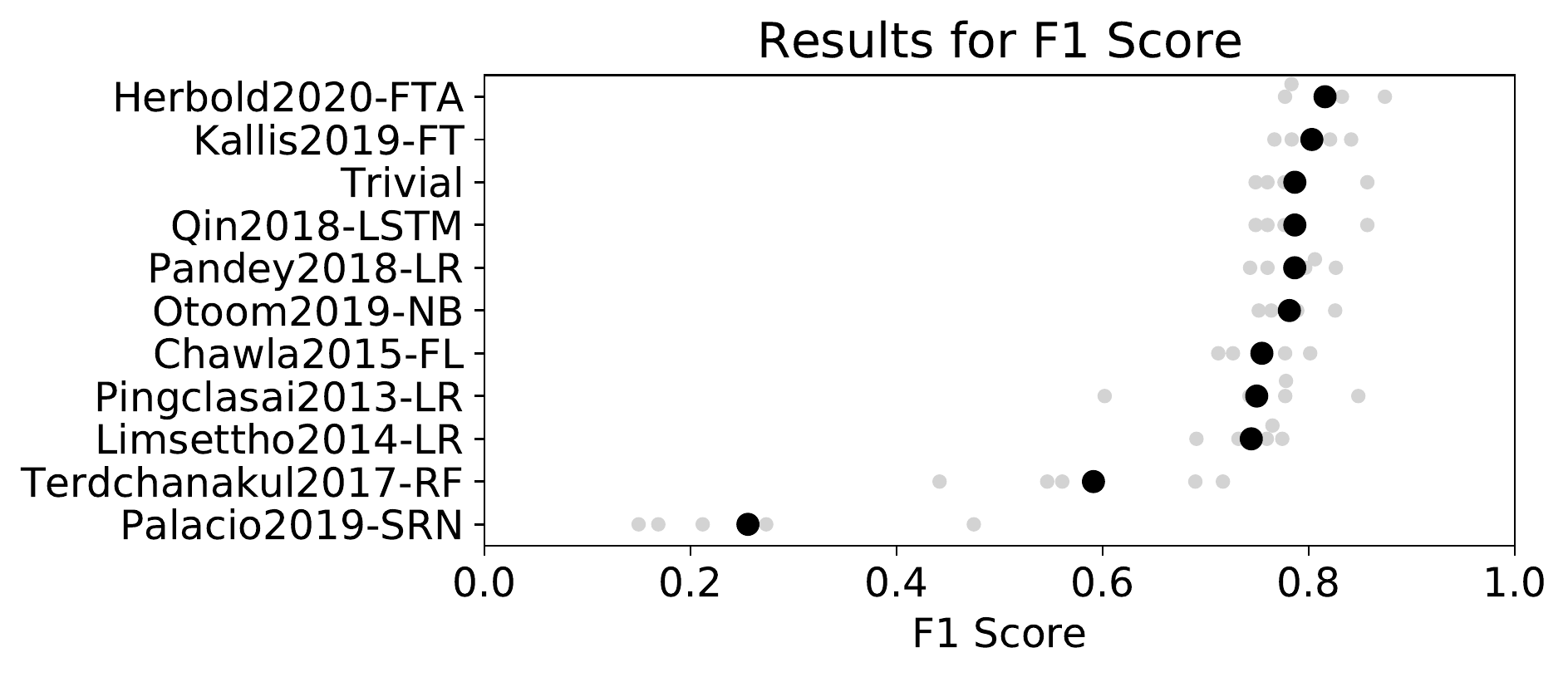}
\end{minipage}
\begin{minipage}{0.49\textwidth}
\includegraphics[width=\textwidth]{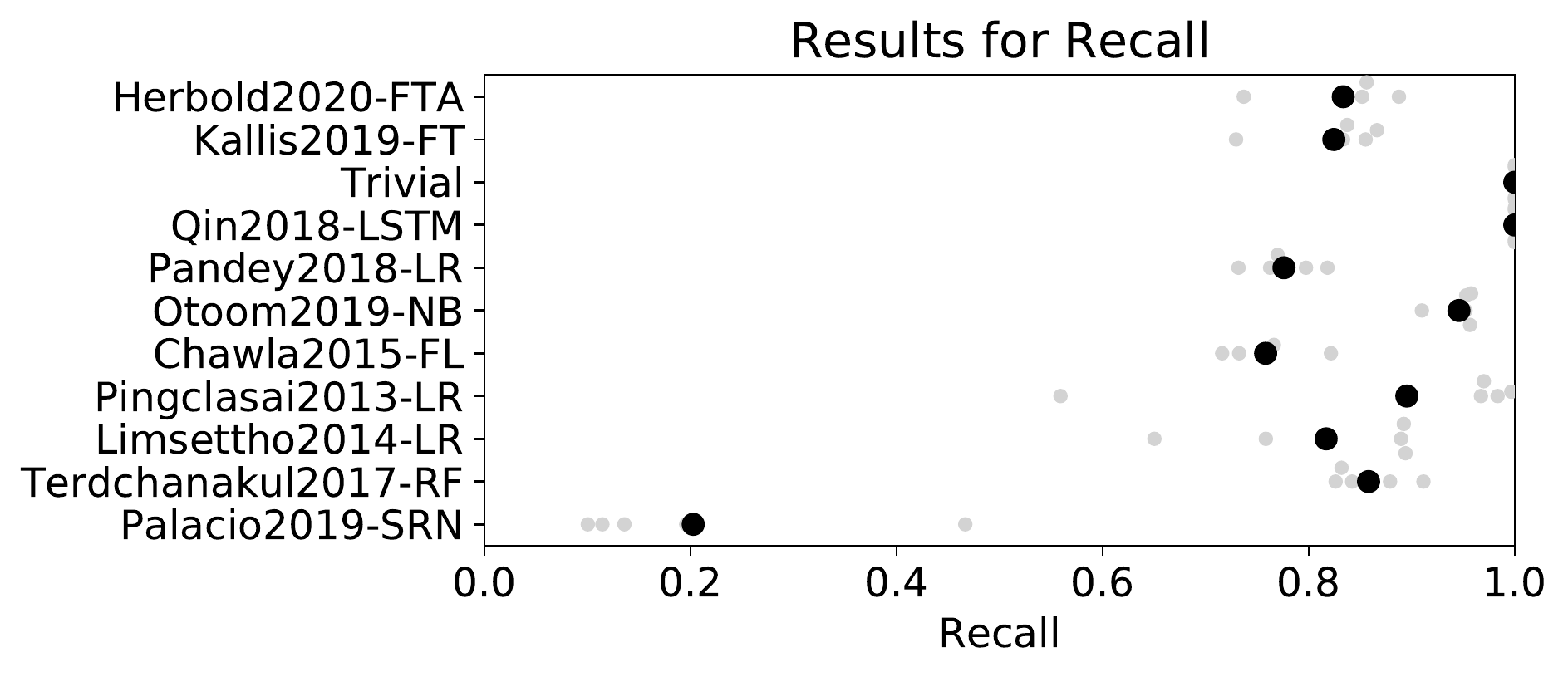}
\includegraphics[width=\textwidth]{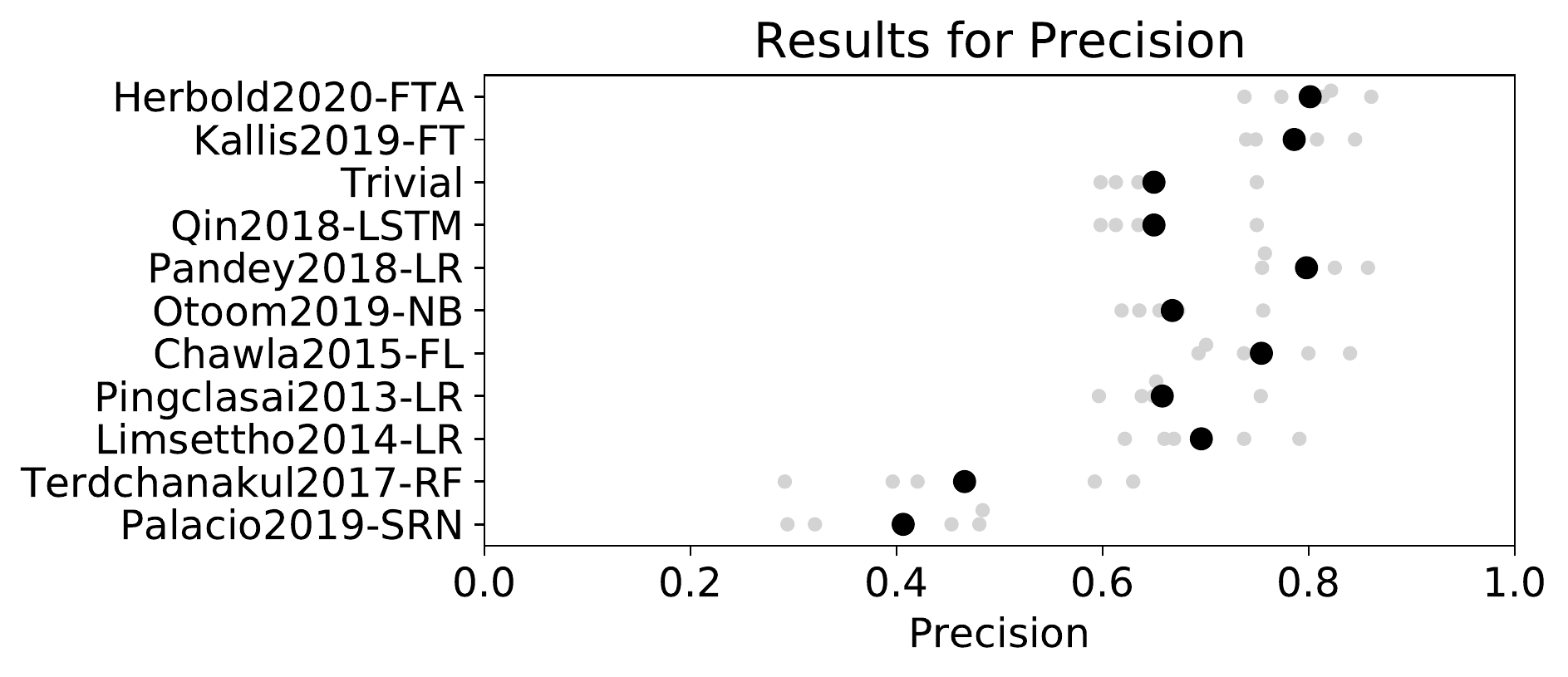}
\end{minipage}
\caption{Results of training with the CV$_{BUG}$ and testing with the TEST$_{BUG}$ data.}
\label{fig:phase2_bugs_only}
\end{figure*}

Figures~\ref{fig:phase2_all_issues} and~\ref{fig:phase2_bugs_only} show the results for the second phase of the experiment, i.e., training with the CV$_{ALL}$, resp. CV$_{BUG}$ data and testing with the TEST$_{ALL}$, resp. TEST$_{BUG}$ data. The results are consistent with our results from Phase 1. Herbold2020-FTA and Kallis2019-FT perform best on both data sets, there is almost no difference on the TEST$_{ALL}$ data and a slightly better \FSCORE{} for Herbold2020-FTA on the TEST$_{BUG}$ data. The mean \FSCORE{} of both approaches is within the confidence interval of the results from Phase 1. In general, we find that even though we just have five projects in the TEST data and the data was independently labelled from the CV data, that the \FSCORE{} of 12 out of 18 approaches is within the CI, in the other six cases (three on TEST$_{ALL}$ and TEST$_{CV}$ each) the values on the TEST data are only slightly higher than the upper bound of the confidence interval on the CV data. This slight improvement of approaches on the TEST data can be explained by the fact that these approaches were developed on the TEST data, which should give them an advantage in comparison to the CV data. Overall, these results are a strong indicator that our findings from Phase 1 generalize to a broader population of projects.

\subsubsection{Results of Phase 3}

\begin{figure*}
\centering
\begin{tabular}{lrrlr}
\toprule
{} &  mean &    sd &              CI &  $d$ \\
\midrule
Herbold2020-FTA-UV & 0.662 & 0.083 &  [0.615, 0.709] &   0.000 \\
Herbold2020-FTA    & 0.648 & 0.075 &  [0.606, 0.690] &   0.179 \\
\bottomrule
\end{tabular}
\hspace{0.5cm}
\begin{minipage}{0.49\textwidth}
\includegraphics[width=\textwidth]{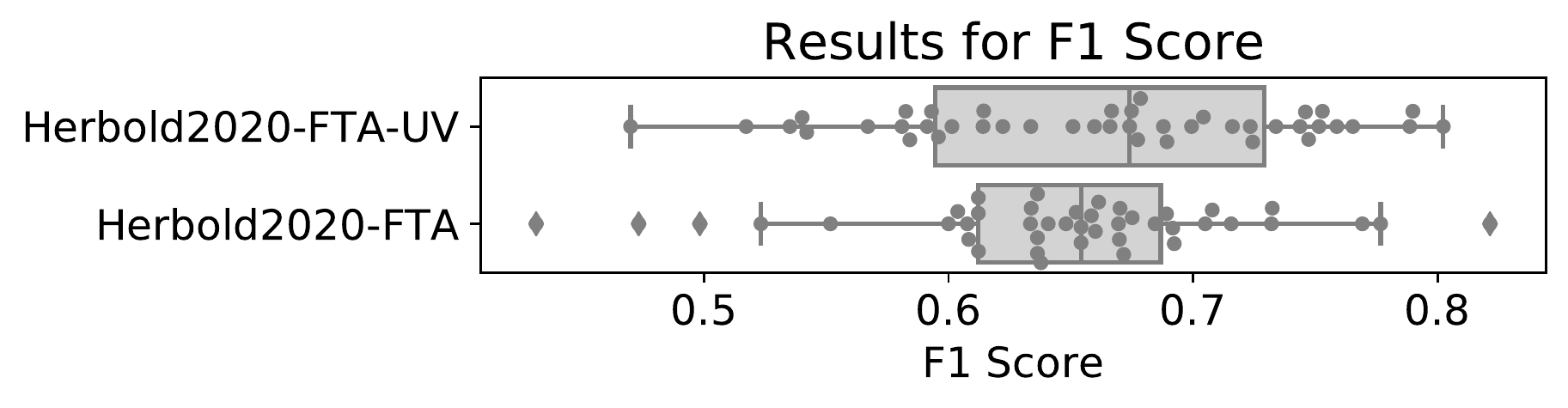}
\end{minipage}
\begin{minipage}{0.49\textwidth}
\includegraphics[width=\textwidth]{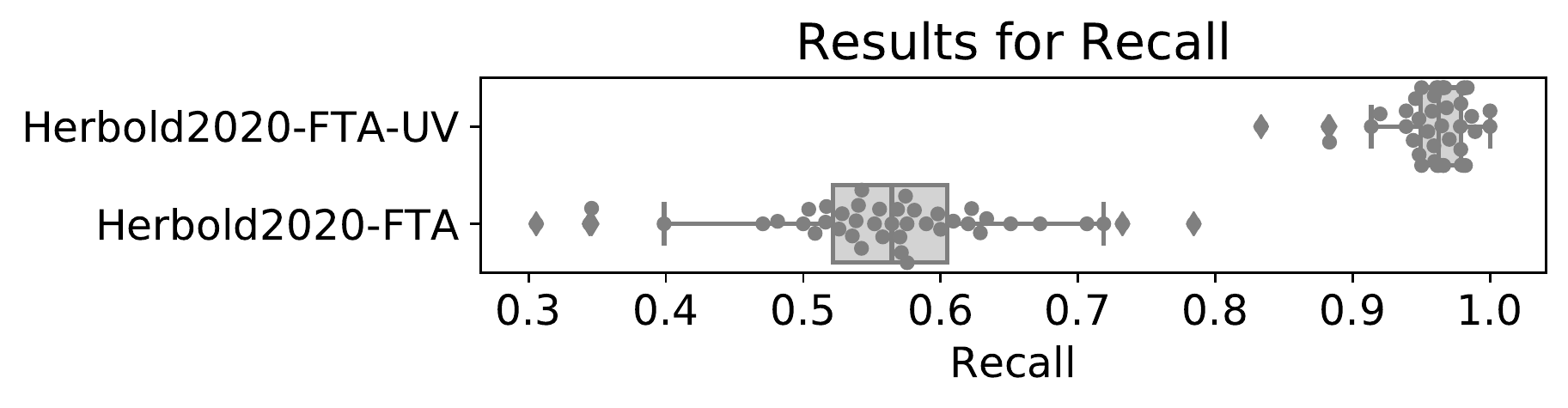}
\includegraphics[width=\textwidth]{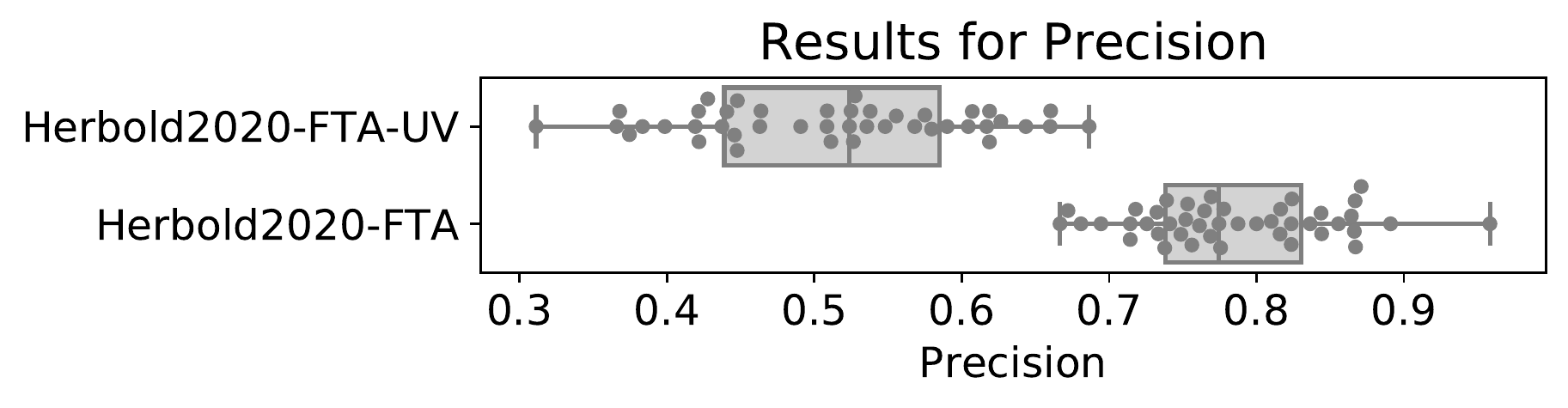}
\end{minipage}
\caption{Comparison of the results from Phase 1 and Phase 2 with training with the UNVALIDATED data and testing with CV$_{ALL}$ and TEST$_{ALL}$.}
\label{fig:phase3_all_issues}
\end{figure*}

\begin{figure*}
\centering
\begin{tabular}{lrrlr}
\toprule
{} &  mean &    sd &              CI &  $d$ \\
\midrule
Herbold2020-FTA    & 0.806 & 0.048 &  [0.779, 0.833] &   0.000 \\
Herbold2020-FTA-UV & 0.783 & 0.060 &  [0.749, 0.817] &   0.435 \\
\bottomrule
\end{tabular}
\hspace{0.5cm}
\begin{minipage}{0.49\textwidth}
\includegraphics[width=\textwidth]{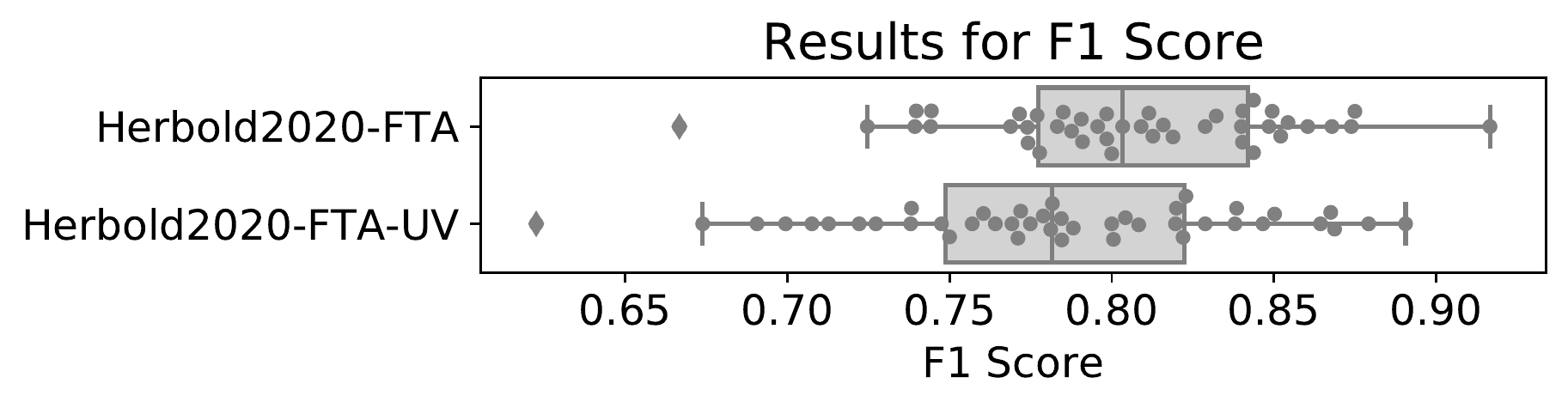}
\end{minipage}
\begin{minipage}{0.49\textwidth}
\includegraphics[width=\textwidth]{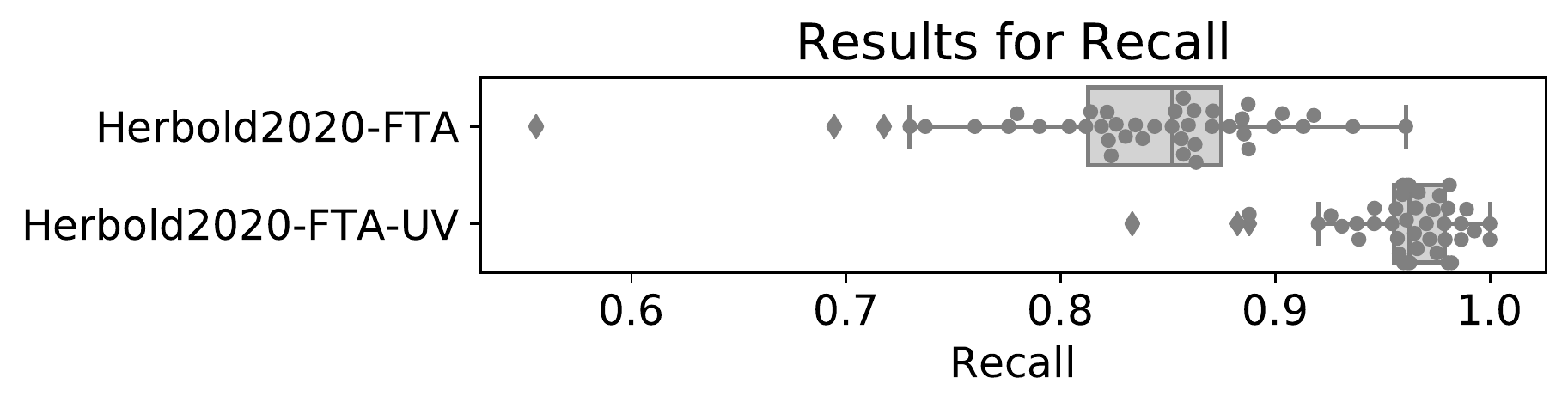}
\includegraphics[width=\textwidth]{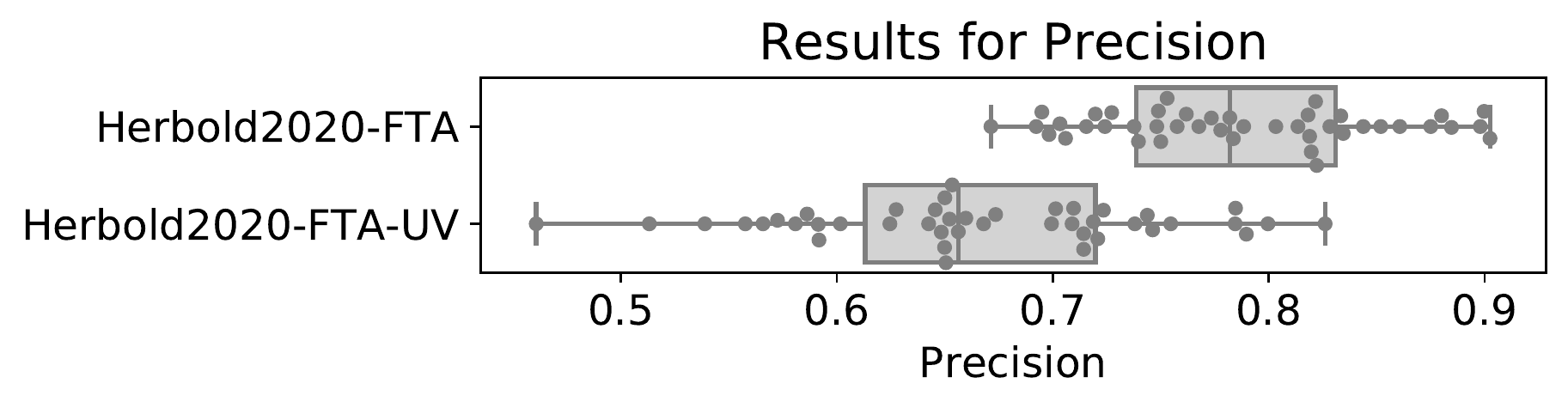}
\end{minipage}
\caption{Comparison of the results from Phase 1 and Phase 2 with training with the UNVALIDATED data and testing with CV$_{BUG}$ and TEST$_{BUG}$.}
\label{fig:phase3_bugs}
\end{figure*}

Figures~\ref{fig:phase3_all_issues} and~\ref{fig:phase3_bugs} show the results of the training with the UNVALIDATED data in comparison to the results of Phase 1 and Phase 2 combined. The training with UNVALIDATED data actually outperforms the training with validated data in case all issues are performed if we consider the mean \FSCORE. However, we fail to reject the null hypothesis of the paired t-test that there is no difference ($p-value=0.385$). Therefore, we conclude that training without manual validation is not significantly different for the identification of bugs among all issues in terms of \FSCORE. Regardless, the \RECALL{} and \PRECISION{} show that while the \FSCORE{} is not affected, the way this \FSCORE{} is achieved is very different between validated and unvalidated data. The Herbold2020-FTA-UV has a very large recall nearly always over 0.9, but a relatively low precision with values between 0.3 and 0.7. Thus, the strong \FSCORE{} is achieved by predicting nearly all bugs at the cost of a mediocre \PRECISION. With the validated data, this is the opposite. The \RECALL{} is much lower and between 0.3 and 0.8, but the \PRECISION{} is higher with values between 0.65 and 1.0. 

If we apply the classifier trained on the UNVALIDATED data to only the bugs, the \FSCORE{} is still similar, but slightly worse than training with validated data. We reject the null hypothesis of the paired t-test that there is no difference ($p-value=0.010$) and find that the effect size of this difference is small ($d=0.435$). In terms of \RECALL{} and \PRECISION{} the differences between training with and without validation are similar. However, the \PRECISION{} of training without validation is increased to values between 0.45 and 0.85 and the \RECALL{} of training with validation is increased to 0.45 to 0.95.
Overall, our results show that training with unvalidated data is an option, especially if \RECALL{} is more important than \PRECISION{}. 

\subsubsection{Results for Phase 4}

\begin{figure}
\centering
\includegraphics[width=0.49\textwidth]{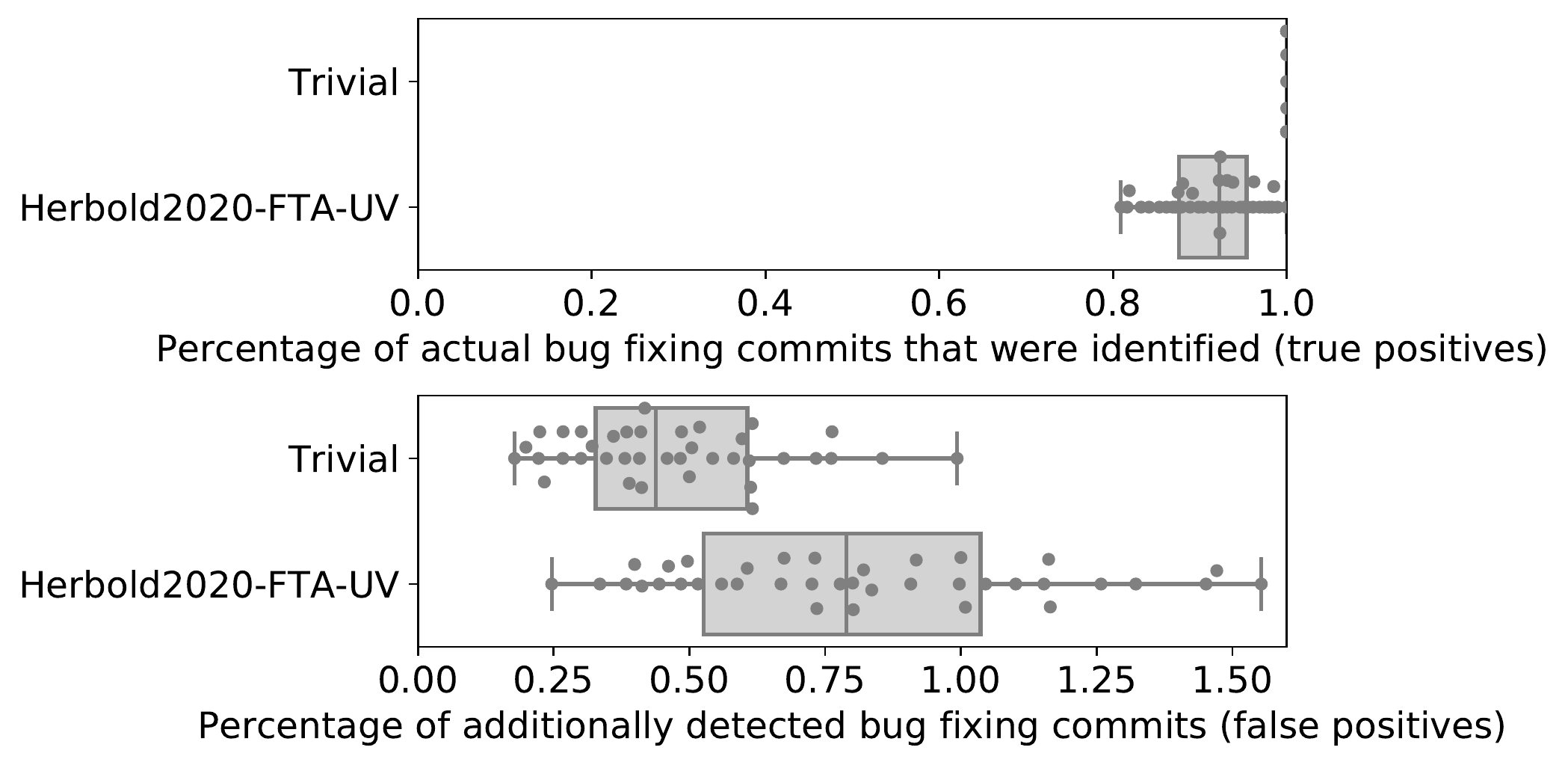}
\caption{Impact on using Herbold2020-FTA-UV for the labeling of bugfixing commits.}
\label{fig:bugfix_labels}
\end{figure}

For Phase 4, we decided to use the Herbold2020-FTA-UV approach, even though Herbold2020-FTA performed slightly better if only bugs are considered. Our reason for this is that we have  a defect prediction use case in mind, where false negatives would mean that bugs are missed. Consequently, we value \RECALL{} higher than \PRECISION{}, which means that Herbold2020-FTA-UV is preferable over Herbold2020-FTA. Figure~\ref{fig:bugfix_labels} shows how the labeling of bugfixing commits is changed by using issue type prediction in comparison to trivially assuming that the developer classification is correct without validation. With the trivial approach, all actual bugfixing commits are identified. This is a property of the data, since the manual validation only reduces the amount of bug fixes. The issue type prediction with Herbold2020-FTA-UV finds on average 91.3\% of the bugfixing commits, the worst case in our data is is that only 80.9\% of the bugfixing commits are identified. When we consider the false positive bugfixing commits, we see that the issue type prediction has a strong positive impact. The mean percentage of additional bugfixing commits is at 47\%, in comparison to 81\% with the trivial approach. Thus, the amount of additional bugfixing commits, that are actually, e.g., feature additions, is greatly reduced by using the issue type prediction. While the resulting data still contains mislabels, the amount of mislabels is reduced. However, even with the reduction there is still much noise left, i.e., about one third of bugfixing commits the data would still be mislabels. Hence, for the use case of the creation of defect prediction data, issue type prediction could potentially replace manual validation for the creation of training data, but we suggest to rely on manual validation for the creation of test data. 

\subsubsection{Results for Phase 5}

Same as for Phase 4, we decided to use the Herbold2020-FTA-UV approach, because this approach performs best when applied to all issues. We use the UNVALIDATED+CV data for training and the CV$_{2014+}$ for testing. This combination of training and test data is realistic for practical scenarios, because there is no temporal overlap between the training and test data. Moreover, the UNVALIDATED+CV data contains historic data from the test projects, that would be readily available without large effort. The advantage of adding this data for the training is that the classifier could possibly also learn project specific information.

\begin{figure*}
\centering
\begin{tabular}{lrrlr}
\toprule
{} &  mean &    sd &              CI &  $d$ \\
\midrule
Time Agnostic & 0.662 & 0.083 &  [0.617, 0.707] &   0.000 \\
Time Aware    & 0.669 & 0.089 &  [0.617, 0.721] &  -0.084 \\
\bottomrule
\end{tabular}
\hspace{0.5cm}
\begin{minipage}{0.49\textwidth}
\includegraphics[width=\textwidth]{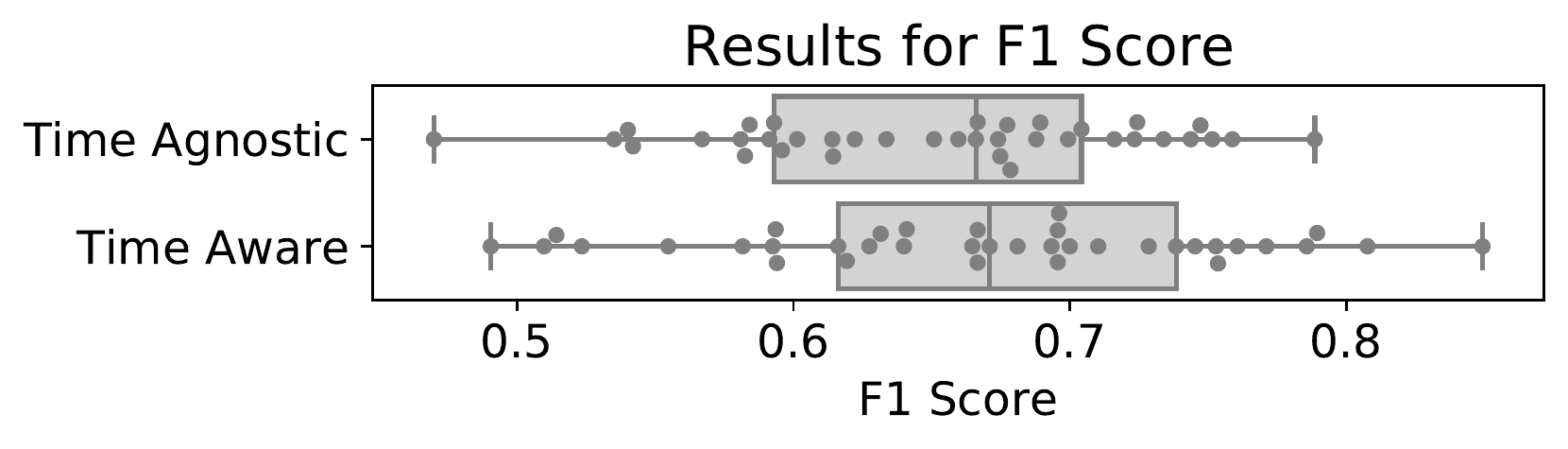}
\end{minipage}
\begin{minipage}{0.49\textwidth}
\includegraphics[width=\textwidth]{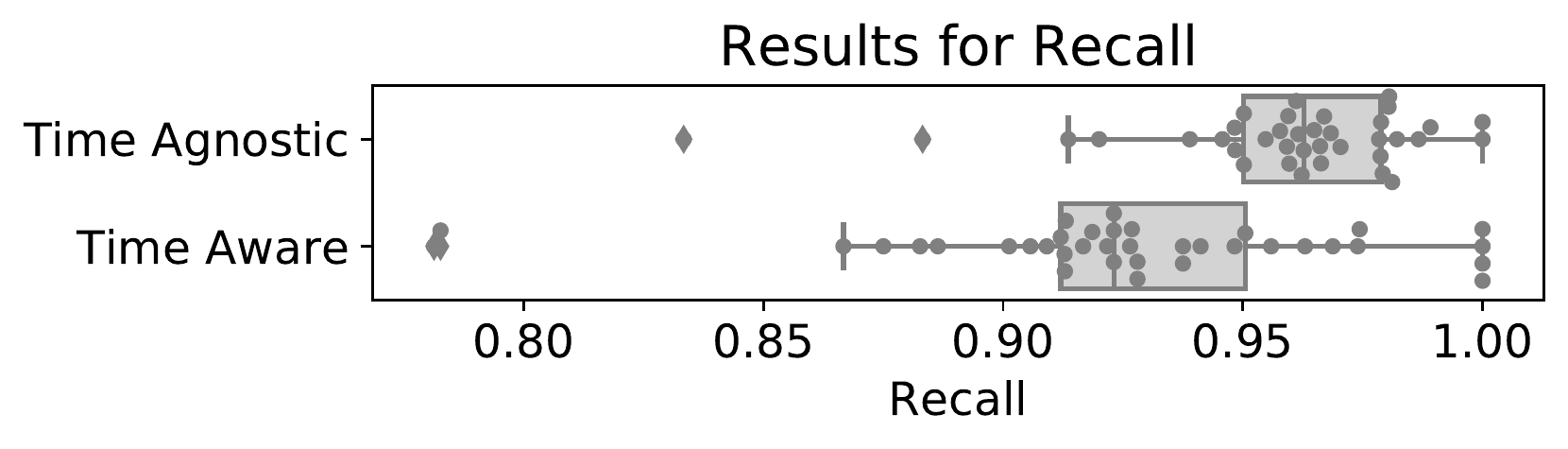}
\includegraphics[width=\textwidth]{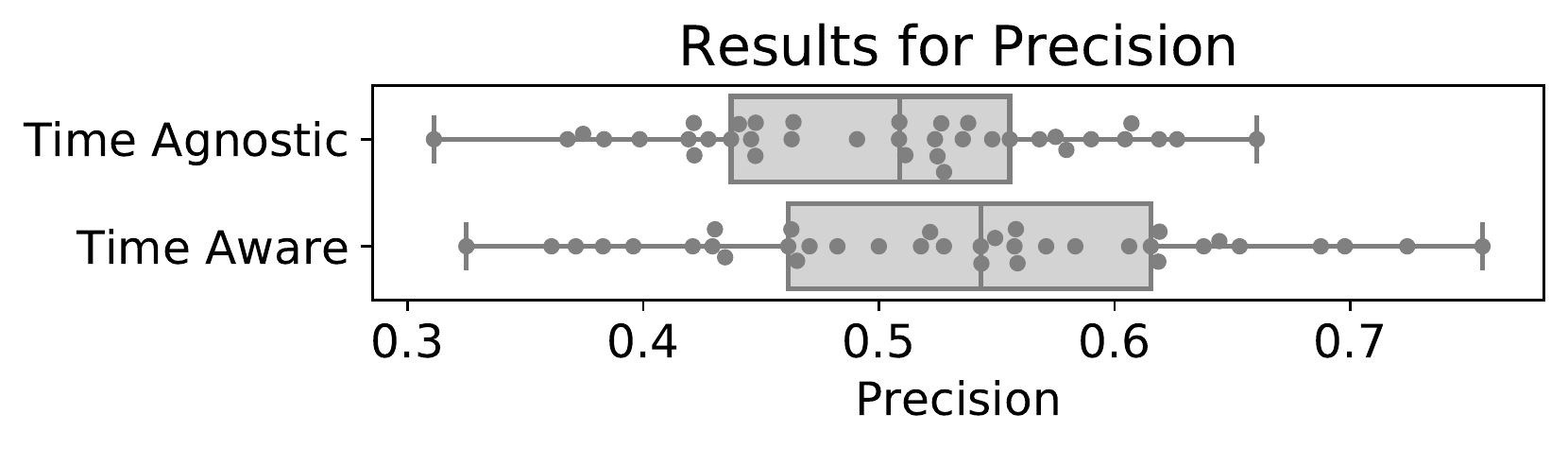}
\end{minipage}
\caption{Evaluation in a realistic time aware setting in comparison to the time agnostic results from Phase 3. The results from Phase 3 are restricted to the projects we evaluate in Phase 5, i.e., the projects from CV without commons-digester.}
\label{fig:phase5_results}
\end{figure*}

Figure~\ref{fig:phase5_results} shows the results of Phase 5. In general, the results are similar to the results of Phase 4. There is a small drop in \RECALL{} and a slight increase in \PRECISION{} with the time awareness, but almost no difference in the F-measure. This is a strong indicator that our results from Phase 3 generalize and that this approach can also be reasonably be used for prediction within issue tracking system. We note that based on the data from \cite{Herzig2013} and \cite{Herbold2020} this prediction system would perform almost the same as developers: most bugs would be correctly labelled as bug (very high recall over 90\%), but only about 55\% of the issues labeled as bug would actually be bugs, which is about 5\% worse than developers.

\section{Discussion}
\label{sec:discussion}

We now discuss the answers to our research questions and how our results relate to findings from the literature.

\subsection{Answers to Research Questions}

\subsubsection*{RQ1: Can manually specified logical rules derived from knowledge about issues be used to improve issue type classification?}

On the one hand, our results show a small improvement in the performance of issue type prediction due to the use of two classifiers, i.e., because of the structural information about the difference between title and description that we incorporate in the learning process. On the other hand, the knowledge about null pointers we integrated to the learning process did not have a positive effect. For us, this indicates two things. First, if we can help the model to better understand the structure of the data, e.g., by accounting for the separate fields, this can have a positive effect, even though the improvement is likely small. Other aspects that could be considered here would be parsing of structural information that is available in the issue, e.g., HTML or Markdown syntax, to enable a better pre-processing of the data. Second, defining logical rules that mimic (parts of) the guidelines for labeling issues by \cite{Herzig2013} does not seem helpful. It seems like the classifiers can infer these rules, or at least similar rules themselves and the definition of hard coded rules may actually inhibit the learning process, because they either restrict the solution space or the available data unnecessarily. In summary, we answer RQ1 as follows. 

\begin{mdframed}
\textbf{Answer to RQ1:} Rules that describe the general structure of issues may improve issue type classification. Rules that describe specific issue types and interfere with the classification should be avoided.
\end{mdframed}

\subsubsection*{RQ2: Does  training  data  have  to  be  manually  validated  or  can  large amount of unvalidated data also lead to good classification models?}

Our results show that training with unvalidated data leads to a comparable mean performance than training with validated data. However, we also find that there are strong differences in how this performance is achieved. Our results provide a strong indication that manual validation leads to models with a better \PRECISION, i.e., fewer false positive predictions of bugs. On the other hand, large amounts of unvalidated data achieve only a mediocre \PRECISION{} that is counterbalanced by a very high recall, i.e., few bugs that are missed. Thus, the choice whether to use unvalidated data or validated data should be done with the use case of the issue type prediction in mind. For example, for defect prediction research few false negatives are important, i.e., unvalidated data is better. If a sample of bugfixing commits should be manually validated line by line to create data like Defects4J~\citep{Just2014}, few false negatives are more important and training with validated data would be preferable. In summary, we answer RQ2 as follows.

\begin{mdframed}
\textbf{Answer to RQ2:} Unvalidated data is useful, if the goal is to miss as few bugs as possible. If there should be few false positive, validated data should be used for training.
\end{mdframed}

\subsubsection*{RQ3:  How  good  are  issue  type  classification  models  at  recognizing  bug issues?}

According to our results, users of issue type prediction can expect a \FSCORE{} of about 0.65 if applied to all issues and of 0.80 if applied to only bugs. Depending on the use case, one can either use validated or unvalidated data and thereby, get usable models for the prediction of bug issues both with few false positives (less than 25\%) or few false negatives (less than 5\%), but not at the same time. However, our results also show the limitations of the current state of the art of issue type prediction. Models that perform universally well with both few false positives and few false negatives are not possible within the current state of the art and the available data. We believe that this could only change, if massive amounts of validated data would be available, i.e., not in the order of $10^5$ as is currently the case, but rather in the order of at least $10^6$ or more. However, given that \cite{Herzig2013} and \cite{Herbold2020} report that the manual issue type classification is very time consuming, it may be problematic to achieve this. In summary, we answer RQ3 as follows.

\begin{mdframed}
\textbf{Answer to RQ3:} Issue type classification models are good at recognizing bugs, in case the use case allows for either some false positives or some false negatives. Current models cannot minimize both false positives and false negatives. 
\end{mdframed}

\subsection{Recommendations for Using Issue Type Prediction}

Based on our results, we have the following recommendations for researchers and practioners with respect to the scenarios we outlined in Section~\ref{sec:terminology}.

\begin{itemize}
\item Researchers may use issue type prediction to reduce the false positive issue labels (Scenario 2), but the resulting data will still contain mislabels and does not constitute ground truth. Models for this purpose should be trained with manually validated data. In case a very high data quality is required, issue type prediction is not a suitable replacement for manual validation. 
\item Practitioners may use issue type prediction to automatically differentiate between bugs and other issue types in bug trackers and get comparable results to the currently assigned labels by developers. Models for this purpose should be trained with a large corpus of data that does not require manual validation. We recommend to use this as a passive recommendation (Scenario 4) and not active recommendation (Scenario 3), because too many wrong active predictions may lead to the rejection of the approach. 
\end{itemize}

\subsection{Comparison with the Literature}

An important part of our work was the replication of the existing approaches from the literature and the evaluation of their performance on independent data and, in general, for more than the usually used five projects by \cite{Herzig2013}. The literature generally reported very good results with performance values higher than the best results we observed in our study. We could not replicate this but are aware of several reasons for the differences between our work and the literature. Most importantly, we used more data. Thus, if an approach randomly works on two or three projects, but fails otherwise, this would be detected by our work, but not necessarily by the smaller case studies in the related work. Also, we used a different case study setup. We clearly separated the training and test data, to avoid any kind of information leakage, e.g., because timing aspects were not considered while doing cross-validation. When we looked at the literature, all prior publications performed some sort of train/test split within the projects, sometimes with accounting for the timing~\citep{Terdchanakul2017,otoom2019automated,Pingclasai2013}, but sometimes not~\citep{limsettho2014comparing,chawla2015automated,qin2018classifying}. In addition to the possible information leakage, this kind of train/test splits reduced the amount of test data to at most 20\% of all the data of a project. Thus, the related work not only used fewer projects, but also less data of these projects for testing. Finally, most of the literature did not even use all five projects from \cite{Herzig2013}, but subsets of this data, further reducing the evidence available for drawing conclusions. Interestingly, the best performing approach from the literature was never evaluated on validated data~\cite{kallis2019ticket}. However, the authors used 30,000 unvalidated issues for the performance estimation, i.e., more evidence than the other approaches from the state of the art. 

In conclusion, we saw that the literature on this topic was not reliable so far and hope that our work not only sheds light on how well issue type prediction actually works, but also serves as guideline for future studies on this topic. 

\section{Threats to Validity}
\label{sec:threats}

There are several threats to the validity of our work, which we report following the classification by \cite{Wohlin2012}.

\subsection{Construct Validity}

The design of experiments may not be suitable to analyze our research question. The biggest threat is to the analysis of RQ1, because we only evaluated the impact of two manually designed rules and only in the single way of using different classifiers. Other manual knowledge or other ways to incorporate this knowledge into the learning process may lead to different results. Moreover, the results of all three research questions are impacted by our choice of \FSCORE, \RECALL, and \PRECISION{} as performance metrics. While these metrics are well accepted in the state of the art and reasonable for the use case, other metrics, especially metrics that would consider different costs for different kinds of misclassification, may lead to different results. 

Regarding the statistical methods, the biggest threat is that we did not perform statistical tests for the selection of the best classifier per publication, but simply took the one with the best mean value. Thus, other classifiers from the same publication might not be statistically significantly different. Unfortunately, performing additional tests was not reasonably possible, because if we would have done one statistical test per publication, we would have a very small significance level for the subsequent tests where we compare publications due to the Bonferroni correction. If we would just have applied the Nemenyi post-hoc test to all 25 approaches, we would have found much fewer significant difference, because the critical distance increases linearly with the number of approaches, i.e., our tests would not have been very powerful. Therefore, doing a selection just with the mean values was the only viable option with the amount of data we have available. Another potential threat is that a Bayesian approach for the statistical analysis as was suggested by \cite{Benavoli2017} as an update of the guidelines by \cite{Demsar2006} may lead to different results. We mitigated this threat by not using pure null hypothesis testing, but also considering the confidence intervals and the effect sizes.

\subsection{Internal Validity}

Our interpretation that the Pandey2018-LR is actually worse than\linebreak Herbold2020-FTA and Kallis2019-FT is only conjectured from the properties of the statistical tests, but not directly supported by the statistical analysis. 
Moreover, our conclusions regarding the differences between the models trained with unvalidated and validate data may also be wrong. The unvalidated data contains more issues, i.e., the size of the data set could also be responsible for the differences in \RECALL{} and \PRECISION{}. We believe that the difference in \RECALL{} may be due to the size and decrease with more validated data, but cannot reasonably determine this without more validated data. However, we believe that the differences in \PRECISION{} are unlikely to go away, because this would mean a performance degradation due to more validated data, which is unlikely.

%Moreover, our determination that the unvalidated data leads to models with high \RECALL{} may also be wrong. The reason for the high \RECALL{} could also be that large amount of data and we would see the same effect if we had over 600,000 validated issues. 

\subsection{External Validity}

The manually validated data was mostly for Java projects of the Apache Software Foundation that use Jira as issue tracker. Only two projects used Bugzilla and one project was from the Mozilla Foundation. The unvalidated data we used was from a diverse range of software projects written in different languages and owned by different organizations, but also limited to Jira as issue tracker. Therefore, it is unclear how well our results generalize beyond Apache projects, the Java programming language, and the Jira issue tracker. However, the good results with the unvalidated data indicate that generalization to different organizations and programming languages is likely. Generalization to other issue trackers is also likely, because we do not observe a difference between the two Bugzilla projects and the Jira projects in our results. Moreover, the results from Phase 2 of our experiments show that our observations hold on two data sets, i.e., at least generalize to some degree. 

\subsection{Reliability}

There are no threats to the reliability of our research, other than the threats to the reliability of the prior research by \cite{Herzig2013} and \cite{Herbold2020}, i.e., the reliability of the manually validated data that we used. However, we note that the work by \cite{Herbold2020} is a successful independent conceptual replication of the work by \cite{Herzig2013}. 

\section{Conclusion}
\label{sec:conclusion}

Issue tracking systems play a central role in the organization of modern software development. Issues that are raised guide the development process and describe, e.g., requested features or reported bugs. Each issue has a type that is assigned by the reporter of the issue and only seldom changed afterwards. From prior research, we know that the issue type does often not match with the content of the issue, e.g., because feature requests are incorrectly labelled as bug. Within this article, we analyzed the state of the art of automated issue type prediction with machine learning and focused on the prediction of whether an issue describes a bug or not. We analyzed if we can improve issue type prediction with rules that account for structural information or rules about whether an issue is a bug and found that accounting for the structure of issues may slightly improve predictions. Moreover, we determined that data that contains mislabeled issues may still leads to good prediction models regardless, especially with respect to their ability to correctly identify all bugs. In comparison, training with manually validated data that does not contain mislabels leads to classifiers with a similar performance, but with fewer non bug issues predicted as bug at the cost of missing more bugs. Overall, the performance of the prediction models is promising and indicates that issue type prediction is likely suitable for practical use within tools and to improve data for research. We demonstrated this with two practical examples that show how the identification of bug fixing commits can be improved and that prediction in issue trackers would be comparable to the labels of developers. 

The knowledge about issue type prediction is still limited, especially regarding the use as recommendation system. Therefore, we believe that future work should investigate how tools such as the issue type prediction tool by \cite{kallis2019ticket} are received by developers and how they can best be integrated into existing issue tracking systems. We believe that such studies should combine quantitative and qualitative aspects, e.g., to quantitatively analyze how often developers agree with predicted labels and qualitatively analyze the feedback of developers through interviews and questionnaires. The development of tools should also consider when and how prediction models are updated. For example, semi-supervised online learning could be used to continuously improve prediction models through re-training in case users actively disagree with predictions. Moreover, we believe that more validated data is the only reasonable way to overcome the problem that one must choose between models with high \RECALL{} and models with high \PRECISION. A collaboration by many researchers with the research turk approach~\citep{Herbold2020b} could help to overcome the issue of the large manual effort and be suitable for the creation of a larger data set. 

\section*{Acknowledgments}

This work is partially funded by DFG Grant 402774445.

\bibliography{./literature}
\end{document}